\documentclass[final,5p,times,twocolumn]{elsarticle}
\usepackage{url}

\usepackage{graphicx}
\usepackage{amssymb}
 \usepackage{lineno}

\usepackage[utf8]{inputenc} %Codificacion utf-8

\biboptions{authoryear}

\usepackage{color}

\journal{Astronomy \& Computing}

\begin{document}

\begin{frontmatter}


\title{The PAU Survey: Operation and orchestration of multi-band survey data}



\author[ifae,pic]{Nadia Tonello\corref{sci_author}}
\author[ciemat,pic]{Pau Tallada\corref{tech_author}}
\ead{tallada@pic.es}
\author[ice,ieec]{Santiago Serrano}

\author[ifae,pic]{Jorge Carretero}
\author[ifae,pic]{Martin Eriksen}
\author[ice,ieec]{Martin Folger}
\author[ifae,pic]{Christian Neissner}
\author[ciemat]{Ignacio Sevilla-Noarbe}

\author[ice,ieec]{Francisco J. Castander}
\author[ifae,pic]{Manuel Delfino}
\author[ciemat]{Juan De Vicente}
\author[ifae]{Enrique Fernandez}
\author[ift]{Juan Garcia-Bellido}
\author[ice,ieec]{Enrique Gaztanaga}
\author[ifae]{Cristobal Padilla}
\author[ciemat]{Eusebio Sanchez}
\author[ippa]{Luca Tortorelli}

\cortext[sci_author]{Main scientific author}
\cortext[tech_author]{Main technical author}

\address[ifae]{Institut de F\'isica d’Altes Energies (IFAE), The Barcelona Institute of Science and Technology, Campus UAB, 08193 Bellaterra (Barcelona), Spain}
\address[ciemat]{Centro de Investigaciones Energ\'eticas, Medioambientales y Tecnol\'ogicas (CIEMAT), Avenida Complutense 40, 28040 Madrid, Spain}
\address[ice]{Institute of Space Sciences (ICE, CSIC), Campus UAB, Carrer de Can Magrans, s/n, 08193 Barcelona, Spain}
\address[ieec]{Institut d'Estudis Espacials de Catalunya (IEEC), E-08034 Barcelona, Spain}
\address[ift]{Instituto de F\'isica Te\'orica, Universidad Aut\'onoma de Madrid, Cantoblanco 28049 Madrid, Spain}
\address[ippa]{Institute for Particle Physics and Astrophysics, ETH Zürich, Wolfgang-Pauli-Str. 27, 8093 Zürich, Switzerland}

\fntext[pic]{also at Port d'Informació Científica (PIC), Campus UAB, C. Albareda s/n, 08193 Bellaterra (Cerdanyola del Vallès), Spain }

\begin{abstract}
The Physics of the Accelerating Universe (PAU) Survey is an international project for the study of cosmological parameters associated with Dark Energy. PAU's 18-CCD camera (PAUCam), installed at the prime focus of the William Herschel Telescope at the Roque de los Muchachos Observatory (La Palma, Canary Islands), scans part of the northern sky, to collect low resolution spectral information of millions of galaxies with its unique set of 40 narrow-band filters in the optical range from 450 nm to 850 nm, and a set of 6 standard broad band filters. The PAU data management (PAUdm) team is in charge of treating the data, including data transfer from the observatory to the PAU Survey data center, hosted at Port d'Informaci\'o Cient\'ifica (PIC). PAUdm is also in charge of the storage, data reduction and, finally, of making the results available to the scientific community. We describe the technical solutions adopted to cover different aspects of the PAU Survey data management, from the computing infrastructure to support the operations, to the software tools and web services for the data process orchestration and exploration. In particular we will focus on the PAU database, developed for the coordination of the different PAUdm tasks, and to preserve and guarantee the consistency of data and metadata.

%% To be submitted to Astronomy and Computing:\\
%% https://www.journals.elsevier.com/astronomy-and-computing/
\end{abstract}


\end{frontmatter}

%%\linenumbers
%%\begin{linenumbers}
\section{Introduction}
\label{intro}

The Physics of the Accelerating Universe Survey (PAUS\footnote{\url{https://www.pausurvey.org}}, \citet{marti}) observes part of the northern sky for the study of the accelerated expansion rate of the universe. The main scientific contribution will be given by the calculation of the photometric redshift (photo-z) of known galaxy catalogs, with an improved resolution of 0.0035 (1+z) \citep{photoz}, together with the study of spectral features, clustering, intrinsic alignments and galaxy evolution, among other science cases.

The project is governed by a consortium, originally founded by the Spanish institutes Centro de Investigaciones Energ\'eticas, Medioambientales y Tecnol\'ogicas (CIEMAT), Instituto de Fisica 
Te\'orica (IFT), Instituto de Ciencias del Espacio and Instituto d'Estudis Espacials de Catalunya (ICE/IEEC-CSIC), Institut de F\'isica d'Altes Energ\'ies (IFAE) and Port d'Informaci\'o Cient\'ifica (IFAE/PIC), with the later incorporation of several European institutes (Durham University, ETH Zurich, Leiden Observatory, University College London) for its scientific exploitation.

The delivery of science ready data products to the PAUS Collaboration and to the scientific community is responsibility of the PAU Survey data management (PAUdm) team. The main challenges faced in this project for the operation and orchestration of PAUS data are typical of a highly automated system, delivering and processing a considerably high data volume (compared to the one that can be comfortably handled by a single computing machine), to be exploited by a scientific community spread over many institutes in different countries. 

Successful big projects in managing and distributing astronomical data, like the Sloan Digital Sky Survey (SDSS\footnote{\url{https://www.sdss.org/}}), are a clear reference for this work, but are hardly reusable, given the rapid evolution of the technical and software tools which can be applied to data management, in addition to the peculiarities and the size of the project.

The PAU Survey is carried out using an imaging camera, called PAUCam \citep{paucam}, designed and built at the engineering facilities of IFAE, in Barcelona. PAUCam is a community instrument installed at the prime focus of the 4.2-m diameter William Herschel Telescope (WHT) at the Roque de los Muchachos Observatory (La Palma, Spain), since mid 2015. PAUCam is made of 18 4k x 2k CCDs, with a system of 6 broad (u, g, r, i, z, Y) and 40 narrow band optical filters (wavelength range: from 450 nm to 850 nm) for a high-resolution photometric survey. The filters are installed in a set of moving, interchanging trays \citep{trays}. With one of the five narrow-band filter trays positioned in front of the focal plane, each narrow band filter covers one of the 8 inner CCDs, while the outer CCDs are covered by broad band filters. Six trays are fully equipped with a standard broad band filter each.

Each PAUCam focal plane image consists of about 650 MiB of information, which translates into a mean total data volume of 200 GiB for a typical observing night. Each night of observing time, PAUCam data are transferred to the PAUS data center, where it is stored and processed with a specifically designed data reduction software. The specific algorithms running during the PAUS data reduction (image detrending and cleaning, astrometric and photometric calibration) and the ones running for the production of the final catalogs (sources extraction and co-added objects spectra) are explained in \citet{paudm1} and \citet{paudm2} respectively.

The infrastructure selected to run the PAU Survey pipelines and to store the data is the High Throughput Computing (HTC) facility available at the Port d'Informaci\'o Cient\'ifica, primarily functioning as a Worldwide Large Hadron Collider Computing Grid Tier-1 facility, but also giving support to Astrophysics and Cosmology research groups like MAGIC\footnote{\url{https://wwwmagic.mpp.mpg.de/}}, for which PIC is the reference data center, MICE\footnote{\url{http://maia.ice.cat/mice/}}, DES\footnote{\url{https://www.darkenergysurvey.org/}}, the European Space Agency mission \textit{Euclid}\footnote{\url{https://www.euclid-ec.org/}} and the PAU Survey.

The core of the PAUS data management (PAUdm) is the PAUS database (PAUdb), on which all the other services rely for orchestrating the jobs for the reduction of the PAUCam images, storing results and accessing them.

In this paper we present the data management system for the PAU Survey operation and orchestration. The constraints guiding the PAUdm design are detailed in section \ref{s:requirements}. The subsequent sections describe the data management approach to fulfill them, highlighting the unique aspects that were developed for processing and making available the PAU Survey datasets. They are: the short term storage at the observatory and transfer procedure (section \ref{s:LaPalma}), the long term storage at PIC (section \ref{s:archive}), the PAUS database for metadata preservation (section \ref{s:database}), the automated orchestration of the nightly data reduction (section \ref{s:operations}), and the web services for data exploration and distribution (section \ref{s:web_services}). Finally the PAUS data center infrastructure is described in section \ref{s:pic}, followed by conclusions with a short discussion of the learned lessons in section \ref{conclusions}.

\section{PAUS data management operations constraints}
\label{s:requirements}

The limited availability of observation time for the PAU Survey at the WHT and the ambitious scientific goals of the project constrain the PAUdm operations design. Those constraints have been formulated in terms of requirements, addressed here in detail.

\begin{itemize}
\item \textit{Short term storage and data transfer.}

All files produced by PAUCam at the WHT shall be transferred to PIC after data taking. A temporal storage in the observatory for a minimum of 5 days must be guaranteed in case of connection failure or other temporary problem.

Data is transferred to PIC the morning after a night of observation. The data buffer at the observatory and the transfer procedure from La Palma are described in section \ref{s:LaPalma}.

\item \textit{Data files preservation.}

The PAU Survey's raw and reduced data files shall be archived and preserved in an organized way. 

While the storage at the observatory acts as a buffer for several days worth of data, the PAU data center shall guarantee the long-term storage of all data sets. The archive system of PAUdm is described in section \ref{s:archive}.

\item \textit{Metadata integrity and preservation.}

PAUdm shall guarantee the preservation and consistency of the PAU Survey's files metadata and of the results of the images reduction, such that they can be accessible for their scientific exploitation.

PAU Survey metadata is checked and preserved in a PostgreSQL\footnote{\url{https://www.postgresql.org/}} relational database, called PAUdb. The organization of the raw metadata, of the reduction results and of the parameters describing how they have been obtained is explained in section \ref{s:database}.

\item \textit{Nightly results availability before the next observing night.}

The pipeline for the image reduction (the \textit{nightly} pipeline) shall deliver a nightly report with the quality of the data in about 6 hours. The survey plan of each observation night is guided by the nightly pipeline results.

The nightly data process orchestration relies on the metadata stored in PAUdb and is described in section \ref{s:orchestration}. The reports system is done querying PAUdb and it is accessible through a web interface, as described in \ref{ss:nightly_report}.

\item \textit{Data and metadata access, distribution and publication}

The PAU Survey files, both raw and reduced images, as well as the analysis results must be accessible, distributed and published, primarily to the collaborators of the project, and finally to the whole scientific community.

A web interface to PAUdb has been designed to access the data process information and the metadata produced during the image reduction and analysis. PAUdb, especially the tables storing the final catalogs, are accessible by the PAU Science group through this web interface. The tools developed to permit the data access will be described in section \ref{s:web_services}.

\end{itemize}

%--------------------------
\section{PAU short term storage and data transfer}
\label{s:LaPalma}

PAUCam is taking exposures of the night sky at the William Herschel Telescope, at the Roque de los Muchachos Astronomical Observatory in La Palma (Canary Islands). Each exposure is a multi-extension FITS\footnote{\url{https://fits.gsfc.nasa.gov/fits_documentation.html}} file
that has to be transferred to the PAU data center, preserving the information (metadata) collected by the PAUCam control system related to the sky and telescope conditions when the data was taken.
The organization and collection of the information related to the FITS files is a joint effort between the PAUCam and the PAUdm teams. The FITS files, with their header information including field coordinates corresponding to the center of each focal plane and weather conditions, are written by the PAUCam data acquisition system \citep{paucam}.

The data is organized in observation sets. Each observation set is a group of contiguous exposures collected in the same directory and taken in the same telescope configuration. The metadata related to each observation set (date, operator comments, project name, list of files names) is collected in a YAML\footnote{\url{http://yaml.org/}} file by the PAUCam data acquisition system.

At the end of a night of observation, the telescope operator gives the command to start the archiving procedure. Data files (FITS and YAML) are moved atomically, together with files attributes, like the adler32 checksum, from a PAUCam disk to the PAUdm disk space, located at the observatory, and functioning as a temporary storage. From this point on, the responsibility of the file passes from the PAUCam team to the PAUdm team. The disk capacity of the temporary storage is 8 TiB, enough to guarantee one week of raw data storage in optimistic weather conditions.

The new observation sets located in the PAUdm disk are transferred to PAUS archive at PIC, using a procedure triggered automatically from PIC every morning. The file transfer consists of two phases: data transfer and \textit{register} job submission (see section \ref{s:dataflow}).

The temporary storage is manually freed once the transfer at PIC archive and the register of data into the PAUS database and nightly data reduction have ended successfully, showing neither data corruption nor inconsistencies with respect to the observation set YAML file content. 

The data transfer is carried out through the use of the bbcp\footnote{\url{https://www.slac.stanford.edu/~abh/bbcp/}} copying tool. This tool maximizes the bandwidth usage even when using a high-latency wide area network (WAN) link, such as in our case, through a 2000 kilometer-long 1 Gb network link. At the end of this phase, all the observation sets at the WHT are synchronized with the data present in the PAUS archive at PIC.

After the data transfer, a \textit{register} job is submitted for execution (see section \ref{s:dataflow}), which takes care of looking up the new observation sets and submitting custom \textit{register} jobs to insert the metadata from each exposure file into PAUdb.

\begin{figure}[!ht]
\centering
\includegraphics[width=1\columnwidth]{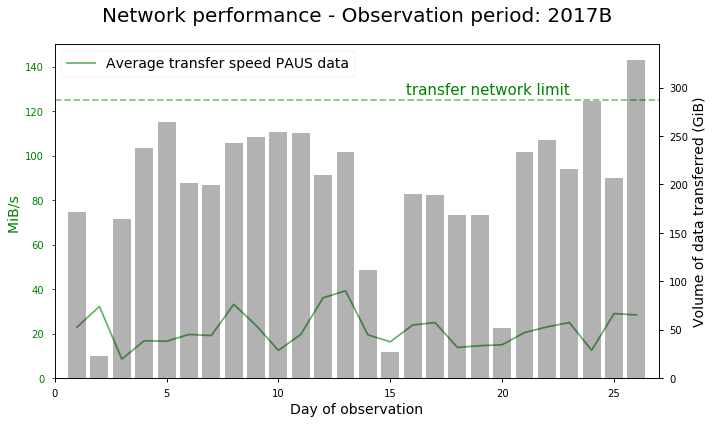}
\caption{The solid line shows the daily average transfer speed registered at PIC for PAU Survey data (observation period 2017B). The histogram shows the data volume transferred in the same day. The transfer rate performance looks quite stable and independent of the transferred data volume. Fluctuations are due to the fact that the network link is shared with other projects of the La Palma observatory.}\label{fig_network}
\end{figure}

Figure \ref{fig_network} shows a sample of the network traffic during the observation period corresponding to the 2017 second semester (2017B), compared with the volume of data downloaded. The download speed is on average around 20 MiB/s, with a good stability over the days, especially considering that PAU Survey shares the network link from La Palma observatory with all the other experiments running in the same astronomical site.

%--------------------------
\section{Data files preservation: The PAUS archive}
\label{s:archive}

\begin{figure*}[!ht]
\centering
\includegraphics[width=0.6\columnwidth]{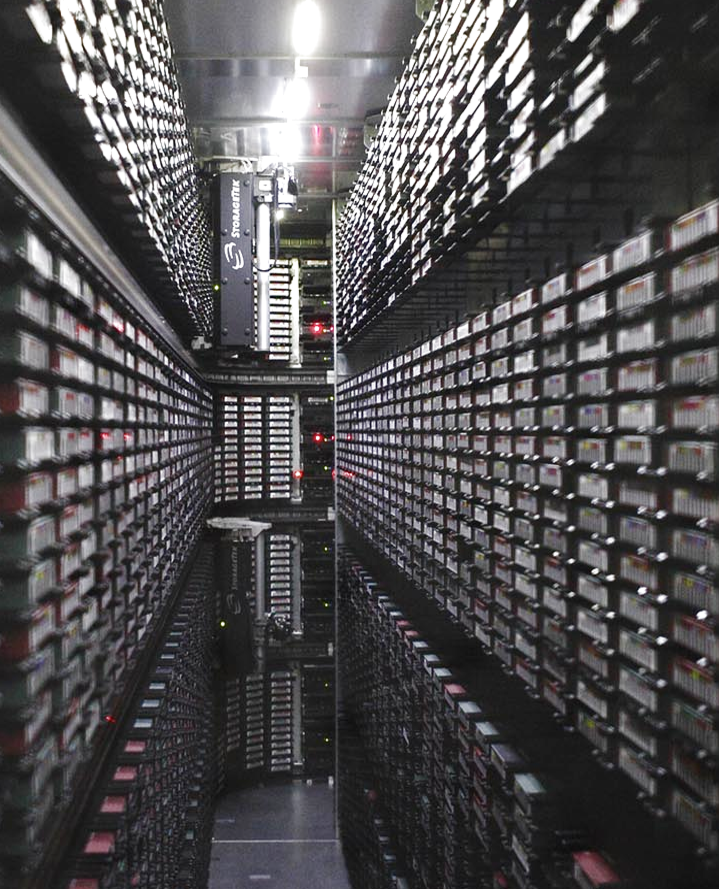}
\includegraphics[width=1.4\columnwidth]{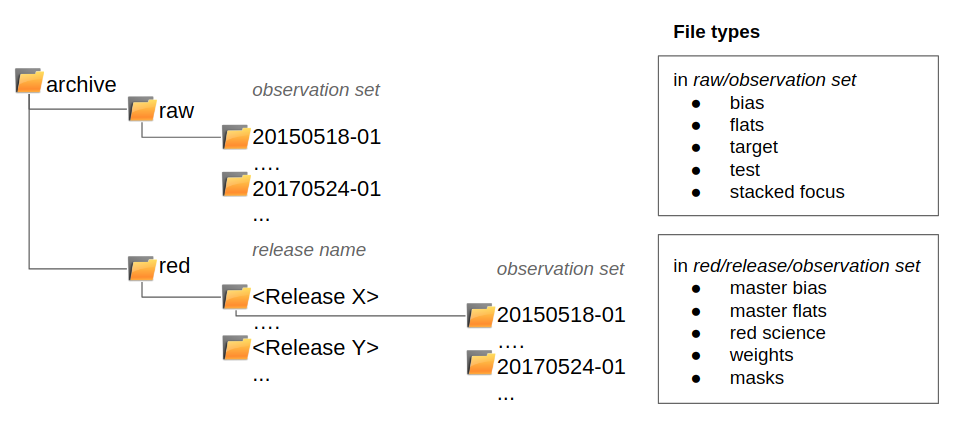}
\caption{Left picture: the Oracle StorageTek SL8500 Modular Library System at PIC, hosting the magnetic tapes where PAUS archive files are stored. Right: PAUS data archive structure along with the different kinds of files it stores. Raw FITS files are organized in folders, each of them identifying a single observation set (see text). Reduced files ('red' folder) are organized per release and observation set.}\label{fig_archive}.  
\end{figure*}

The PAUS archive stores the files produced by the PAU Survey project during normal operations in an organized structure, as shown in Figure \ref{fig_archive}.

The raw data produced by PAUCam is saved in multi-extension FITS (MEF) files called mosaics, corresponding to one focal plane composed by the 18 PAUCam CCDs images. Each extension contains data produced by the readout of one of the 4 amplifier of each CCD, for a total of 72 extensions.

As mentioned in the previous section, a group of FITS files coming from an uninterrupted session of PAUCam observation activity (ideally one per night) is contained in one folder and constitutes an observation set, identified by date of creation and a counter, which is reset at the beginning of each night of observations. Each observation set collects several types of raw files (test files, calibration files, science images). Raw files are classified by their content in different types: bias, flats, sky images, stacked focus and tests. Flats and sky images are classified also based on the filter tray in front of the focal plane when the exposure was taken.

Raw data is processed by the \textit{nightly} pipeline (see section \ref{s:dataflow}) and the resulting products are reduced FITS files. Master bias and master flats files are stored in MEF files, with one extension containing data of one CCD (after amplifiers over-scan and gain correction). Sky images are still stored in FITS files, but data of each CCD is stored separately: reduced (clean) data, weights and corresponding mask. It reduces the data I/O as the data unit of the \textit{memba} pipeline is one CCD, while in the \textit{nightly} pipeline is a whole mosaic.

Each time data is processed with a different code release, the resulting reduced files are stored in a separate folder, whose name corresponds to the release given name.

\begin{figure}[!ht]
\centering
\includegraphics[width=\columnwidth]{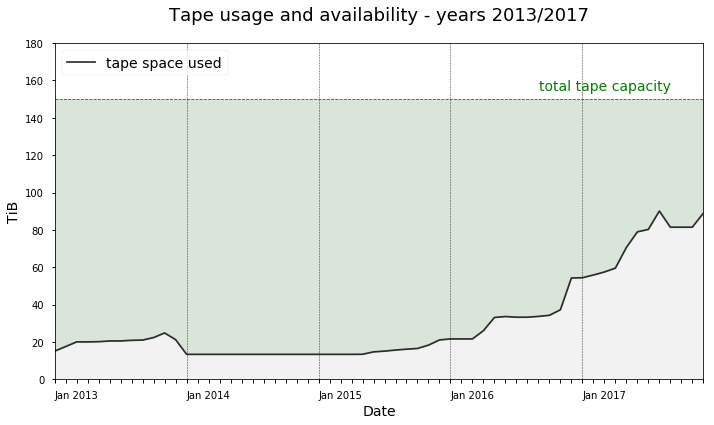}
\caption{Storage space assigned to the PAU Survey project at PIC and its usage between the years 2013 and 2017. Before June 2015, corresponding to PAUCam commissioning, the space has been occupied by simulations and results of the test pipelines prototypes over them. From June 2015 to the present, the storage has been filled with observed and reduced data.}\label{fig_storage}
\end{figure}

The total volume occupied by PAUS files (simulated, raw and reduced) from years 2013 to 2017 is shown in Figure {\ref{fig_storage}}. The planned data volume of 150 TiB has been estimated for the total lifetime of the project and for the most optimistic scenario of the telescope activity from the design phase. 

PAUS data are permanently stored in magnetic tapes. This type of support, optimal for low access rate data, such as actual PAUS data, gives multiple advantages with respect to traditional spinning disks: the reduced cost of the cartridges, the small physical space occupied per TiB, and the limited energy consumption for maintenance are some of them. The main disadvantage is the access latency due to the fact that the volume has to be mounted and sought to access the files (read and/or written), operation that can take around one minute\footnote{Nominal values are: load 13s, file access 50s, unload 23s.}, while the read and write operation from and to tape has a nominal efficiency of 250 MiB/s (dependent on the files size).
The organization of the files archive explained above has been designed taking into account the storage support. The nightly and the analysis operations are typically carried out per groups of observation sets, or per code release respectively. By organizing the data of the same directory in the same cartridge, the number of loads are minimized, for a longer lifetime of the support. The process of load and file access is automated and transparent to the user. For more details about the magnetic tapes technology handled at PIC, see Appendix A.    

The number of tapes assigned to the PAUS projects has changed in the years due to the change in tape technology, but the mean assigned space has been proven to be reasonable.

Data ownership and security has been guaranteed with the creation of a dedicated group of PIC users (paus) and with restricted write permissions to administrators (through dCache ACLs) on the archive space.

\section{Meta-data integrity and preservation: the PAUS database (PAUdb)}
\label{s:database}

The preservation of the data and the metadata, as well as their consistency, is a critical point of the data management of every scientific project. The PAU Survey project is generating a large amount of files while data is being taken and processed, and a large amount of metadata as a result of running image analysis pipelines. The data volume is only one of the critical points: the files and metadata produced at high velocity need to be accessed concurrently by several hundreds of clients, i.e. the nodes of the PIC computer farm where the reduction and analysis code run.

In order to enable the storage, management and distribution of all this information, several alternatives for implementing a metadata repository were taken into account, such as a nested structure of ASCII files, relational databases or some newer NoSQL solutions, taking into account the data management and final users needs. In particular, we looked for a solution allowing for the use of SQL query language, which is already familiar to most of the scientific community, a relational database solution, to allow for comparisons between datasets, and finally a free, mature and stable software, capable to deal with the data volume we had foreseen for the project.

We settled on a relational database setup consisting of two twin servers configured one as a replica of the other. Each server has 12 physical cores, 96 GiB of memory and 2 TiB of storage. We selected PostgreSQL as the actual software solution for the relational database due to its stability, performance and broad compatibility with current SQL standards.

The PAU processing pipelines make heavy use of database transactions to ensure the integrity and consistency of the ingested data. The information stored in the PAU database is backed up periodically to minimize data loss in case of malfunction or catastrophic failure. Critical tables are backed up once a week, while others just once a month. For those tables that are not modified, such as external catalogs, they are dumped only once when they are ingested for the first time.
  
\subsection{PAUdb data model}

\begin{figure*}[!ht]
\centering
\includegraphics[width=1.8\columnwidth]{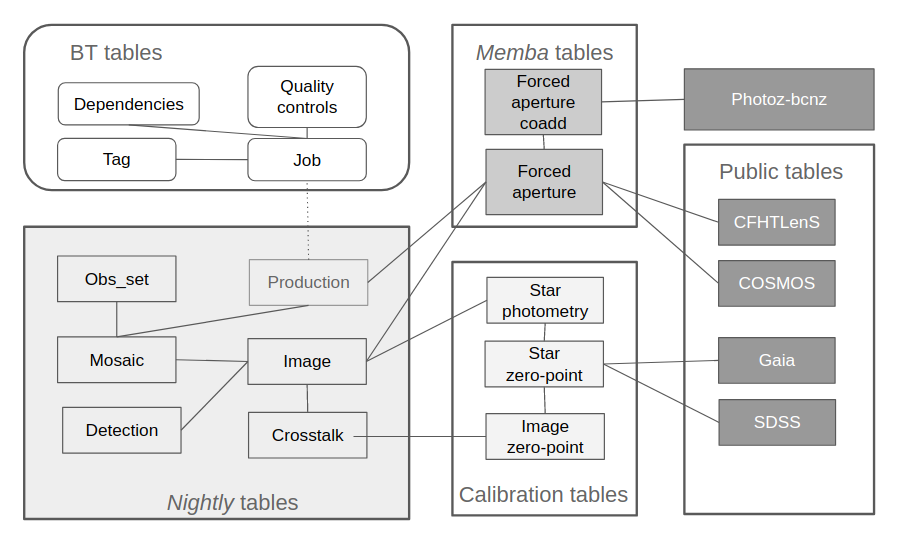}
\caption{PAUdb tables, schematic organization of the main PAUdb tables. For description and list of columns see tables 1, 2, 3, 4 and 5 in Appendix B.}
\label{fig_paudbschema}
\end{figure*}

The PAUdb content is organized in about 40 different tables. Figure \ref{fig_paudbschema} shows the high level organization of the main PAUdb tables (see Appendix B for table description and columns). Some of them are filled and queried during data reduction: they contain the metadata of raw files and files produced by the \textit{nightly} pipeline, as well as calibration factors and intermediate values. The \textit{production} table takes care of the code version preservation. BT (see \ref{s:orchestration}) tables store configuration and metadata related to the tasks run in the computer farm, connected to the tables storing quality checks performed during the data reduction. Other tables store the results of the multi-epoch multi band process intermediate and final catalogs. Public catalogs of reference surveys (Gaia\footnote{\url{https://cdn.gea.esac.esa.int/Gaia/}}, SDSS\footnote{\url{https://skyserver.sdss.org/dr14/en/home.aspx}}, CFHTLenS\footnote{\url{http://www.cadc-ccda.hia-iha.nrc-cnrc.gc.ca/en/cfht/}}, etc.) are stored in PAUdb for calibration and analysis purposes. 

PAUdb, like all relational databases, is interfaced using Structured Query Language (SQL)  commands or statements. This language enables users to specify, in a declarative manner, the operation they want to perform on the database, such as retrieving a subset of rows or modifying some existing values. Writing complex SQL statements usually requires full knowledge of the database model and understanding of how relational databases work. There are also strong security concerns when those statements contain user supplied input, as that may lead to unexpected results such as information leak, alteration or destruction.

In order to mitigate all those issues and to facilitate the access to PAUdb by the pipeline developers, we decided to proxy all database operations through an Object Relational Mapper (ORM), allowing users to interface with the database using the standard constructions present in their programming language. Having Python\footnote{\url{https://www.python.org/}} as the main programming language for the PAUS processing pipelines, we chose SQLAlchemy\footnote{\url{https://www.sqlalchemy.org/}} as the specific ORM solution because of its complete feature set and its comprehensive documentation.

The database structure, described in the ORM model, allows the developers to access the database by importing a specific Python module in their code and using the  set of classes, objects and methods defined in it, instead of having to manually construct SQL statements. For instance, querying data from another table linked by a foreign key is as easy as accessing a particular object attribute. Using such an abstraction layer comes with additional benefits, such as being able to change and evolve the database structure without interfering with users, as they only interact with the ORM model.

\subsection{Limitations and improvements}

Even though the implemented relational database is working fine for the PAUS reduction pipelines, it is not optimized to handle high metadata volumes produced by analysis jobs. The database volume grows with the number of exposures, implying an increasing of processing time for large scale analysis.

To ease out the analysis tasks and to facilitate the publication and distribution of the data, the biggest tables handling the products of the image reduction and multi-band analysis are migrated to the PIC big data platform. Once there, several services for interactive analysis and distribution, such as CosmoHub (see \ref{s:cosmohub}), make those tasks easier, faster and without limitations in data set size.

\section{Nightly results availability: PAUdm operations}
\label{s:operations}

PAUdm operations are entirely carried out on top of the infrastructure available at PIC, the PAU data center (see section \ref{s:pic}).

The PAU database has dedicated tables, created in order to automatically activate the execution of functions and orchestrate the nightly operations.

The pipelines, written in Python, are coded in order to be run both in a HTC infrastructure or in a local computer. 

The code of PAUdm is organized in pipelines. Each pipeline consists of one or more types of tasks. Each task type is connected to others by static dependencies. Tasks of the same type have a defined configurable set of parameters and can run in parallel to guarantee the scalability with the number of files to treat. Each task is composed of three parts: a \textit{prolog}, a \textit{run} and an \textit{epilog}. The \textit{prolog} and \textit{epilog} of the task are able to generate new sub-jobs, with the associated dependencies and configuration. The \textit{run} part executes the selected functions to the given input file(s).

\begin{figure*}[!ht]
\centering
\includegraphics[width=1.8\columnwidth]{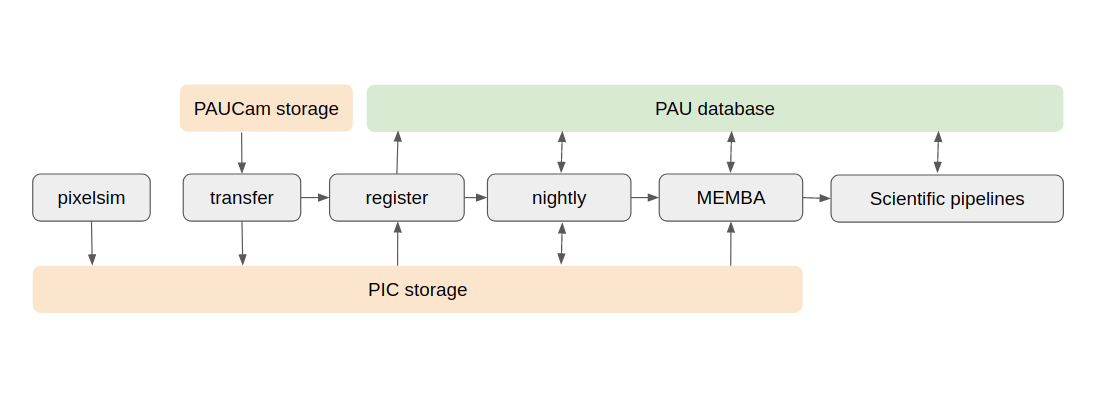}
\caption{PAU data management pipelines dependencies. The white rectangles identify the pipelines and the arrows their interface with the storage system and the PAU database.}\label{fig_dataflow}
\end{figure*}

In PAUdm we have defined the following main pipelines, schematically shown in Figure \ref{fig_dataflow}: \textit{register}, \textit{nightly} and \textit{memba}. Additional pipelines are \textit{pixelsim}, for the simulation of raw PAUCam images, and \textit{crosstalk}, for the evaluation of the crosstalk effect on raw mosaics and its correction. Scientific pipelines (for example, photometric redshift calculation as described in \citet{photoz}) are currently under development from existing separate algorithms, with the intention of integrating them on PAUdm once reaching enough maturity and stability.

In order to execute a task in the PIC computer farm, the functions defined in the run must fit to one job running in one node (defined by the data center as running in 1 core and a top of 3-4 GiB of RAM consumption).

\subsection{Data flow}
\label{s:dataflow}

\begin{figure*}[!ht]
\centering
\includegraphics[width=1.8\columnwidth]{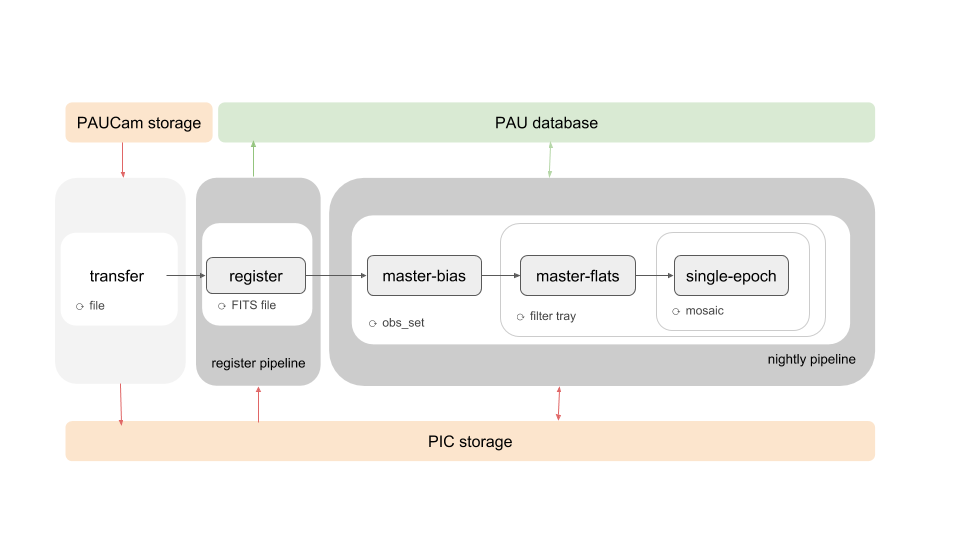}
\caption{PAU data management \textit{transfer}, \textit{register} and \textit{nightly} pipelines, and their relation with storages and the database. The pipeline that registers the mosaics and images in PAUdb is parallelized per FITS file. Jobs in the \textit{nightly} pipeline run following a hierarchical structure: for each observation set, the master bias is calculated first, followed by the master flats. Finally the \textit{single\_epoch} jobs that process each science mosaic is run in a separate job, as soon as the corresponding master-flats mosaic is available.} \label{fig_nightlydataflow}
\end{figure*}

The dependencies and the level of parallelization of the tasks running automatically each night of observation are described schematically in Figure \ref{fig_nightlydataflow}.

\begin{enumerate}

\item After the data has been transferred to PIC, a \textit{register} task per observation set is created by the transfer script. In the \textit{register} task, the \textit{prolog} checks for the list of files belonging to the observation set, as listed in the observation set YAML file, and creates one corresponding \textit{register} subtask job per file. The \textit{register} task reads the header of the FITS file and inserts in PAUdb the metadata corresponding to the mosaic (contained in the exposure file primary header) and the individual images (file extensions headers).

\item The \textit{nightly} pipeline consists of a tree of subtasks that are created per each observation set, whose final result is the production of a master bias (task \textit{master\_bias}, one per observation set), master flat files (task \textit{master\_flats}, one per observation set and per filter tray), reduced images (task \textit{single\_epoch}, one per target file), and the calculation of image parameters and calibration factors that determine the data quality, previous to the \textit{memba} (multi-epoch multi-band) pipeline execution. The algorithms used in the \textit{nightly} pipeline are described in \citet{paudm1}.

The subtasks composing the nightly pipeline (master bias, master flats and single epoch) are orchestrated as follows.
\begin{itemize}
\item Once the metadata of all the FITS files of an observation set is registered in the PAUdb, the \textit{nightly} task for that observation set is created by the epilog of the \textit{register} task.
\item The prolog of the \textit{nightly} task initiates the \textit{master\_bias} task
\item The prolog of the \textit{nightly} task queries PAUdb looking for flats files, creating a \textit{master\_flats} for each kind of flats file found (one master flat per filter tray).
\item If target mosaics are also found, a \textit{single-epoch} subtask is created for each of them. In this phase of the process, the image detrending, the astrometric correction and photometric calibration are performed, taking external public catalogs  as reference for bright stars position and magnitude calibration. 
\end{itemize}

\item While the \textit{register} and the \textit{nightly} pipelines are run the morning after each night of observation, the \textit{memba} pipeline is run periodically, typically after each observation period. In the \textit{memba} pipeline, the flux of stars and galaxies of each image, whose position is taken from a selected photometric catalog (defined in the task configuration) is estimated. Then the information collected in the multiple observations of a single celestial object in different images is added up, and the final output is a catalog with calibrated fluxes of each object in all the observed wavelengths, corresponding to the 40  narrow band filters, and eventually the 6 broad band ones. The algorithms applied to obtain the results are described in \citet{paudm2}.

The \textit{memba} pipeline consists of two types of tasks:
\begin{itemize}
\item The \textit{forced\_phot} task computes the forced aperture flux (using the forced photometry method) of each object per image. The parallelization level is of one job per CCD.
\item The \textit{forced\_phot\_coadd} task co-adds the forced photometry results and obtains a final estimate of flux per object and per waveband. This last task is parallelized in chunks of objects of the selected catalog and operates only at database level.
\end{itemize}
The \textit{memba} pipeline does not create new files, all the intermediate and final results are catalogs stored in dedicated tables of PAUdb.

\end{enumerate}

\subsection{Job orchestration tool: BT}
\label{s:orchestration}

The orchestration of PAUdm jobs is carried out by a specific tool called \textit{brownthrower} (BT). Albeit developed for the PAUdm pipelines, BT is a generic tool that may be used by other projects: in fact, it has been used for orchestrating several MICE and \textit{Euclid} job productions. Developed on top of the PIC grid job scheduler (PBS, HTCondor), it has the advantage of being able to establish and manage the dependencies between different jobs, therefore allowing the implementation of complex pipelines.

The cornerstone of this tool is a relational database, currently in the same database server as PAUdb. It holds all the information about the jobs, such as their input and output data, dependencies, and configuration settings. BT makes heavy use of the transactional nature of the relational database to ensure the consistency and integrity of that information while tracking the status of every job. 

BT provides two tools to manage jobs: a manager and a runner. The BT manager is a command line interface to create, configure and submit jobs for execution, querying their status and abort them. The BT runner is the tool that executes the job. Once launched in the proper environment, it starts pulling ready-to-run jobs from the database and executes them sequentially. In practice BT runner is used as a pilot job, where several hundreds of instances are running all the time in the computing farm, executing many jobs in parallel. Finally, the fact that all the jobs, past and present, are stored in a relational database, makes all data available for auditing and accounting purposes. 
The PAUdm jobs, organized in highly parallelized tasks, are orchestrated thanks to the connection between BT and PAUdb, both for jobs created automatically and manually.

\begin{figure}[!ht]
\centering
\includegraphics[width=1\columnwidth]{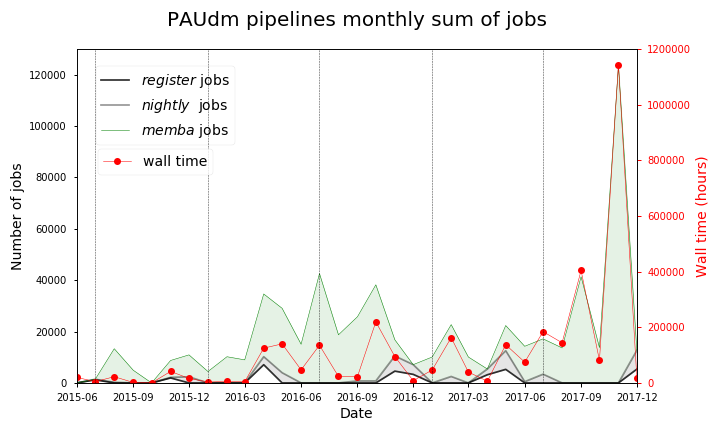}
\caption{Monthly sum of the number of jobs (solid lines) and wall time (red dots), from commissioning on June 2015 to the end of 2017. Only PAU \textit{register}, \textit{nightly} and \textit{memba} jobs have been selected (other analysis and test jobs have been removed for clarity).
Register jobs have been run during the PAUCam observation periods. \textit{Nightly} jobs follow the \textit{register} pattern. Extra \textit{nightly} code releases have been run out of the observation periods over subfields of the survey for validation purposes. Most of the time spent in the computer farm is due to \textit{memba} jobs, the high number of jobs depending on the high level of parallelization of the pipeline. \textit{Memba} jobs have been run with different configuration and different code complexity, explaining the non-linear dependency of the number of jobs from the wall time.}\label{fig_njobs}
\end{figure}

Figure \ref{fig_njobs} shows part of the computer farm activity for PAUS, from the commissioning of PAUCam to the end of 2017, both in terms of number of jobs (each one of them occupying one slot in a node) and on wall time, i.e. the time spent from queue to the job end. 

\section{PAUdm web services for data access, distribution and publication}
\label{s:web_services}

We developed a web based graphical user interface to PAUdb, which provides simple access to many parts of the PAUdm system.

Its functionality, described in the following sections, includes a monitor of job progress (section \ref{ss:operations_control}) and data quality during observation runs (section \ref{ss:data_quality}), the nightly data report summary (section \ref{ss:nightly_report}), access to raw and reduced image data (section \ref{ss:data_access}), a description of the PAUdb schema and an SQL Browser, and an inspector of flux measurements of objects using the 40 PAUS narrow band filters based on the data used to calculate their final co-added values (section \ref{aperture_inspector}).

\subsection{Operations control}
\label{ss:operations_control}

\begin{figure*}[!ht]
\centering
\includegraphics[width=1.5\columnwidth]{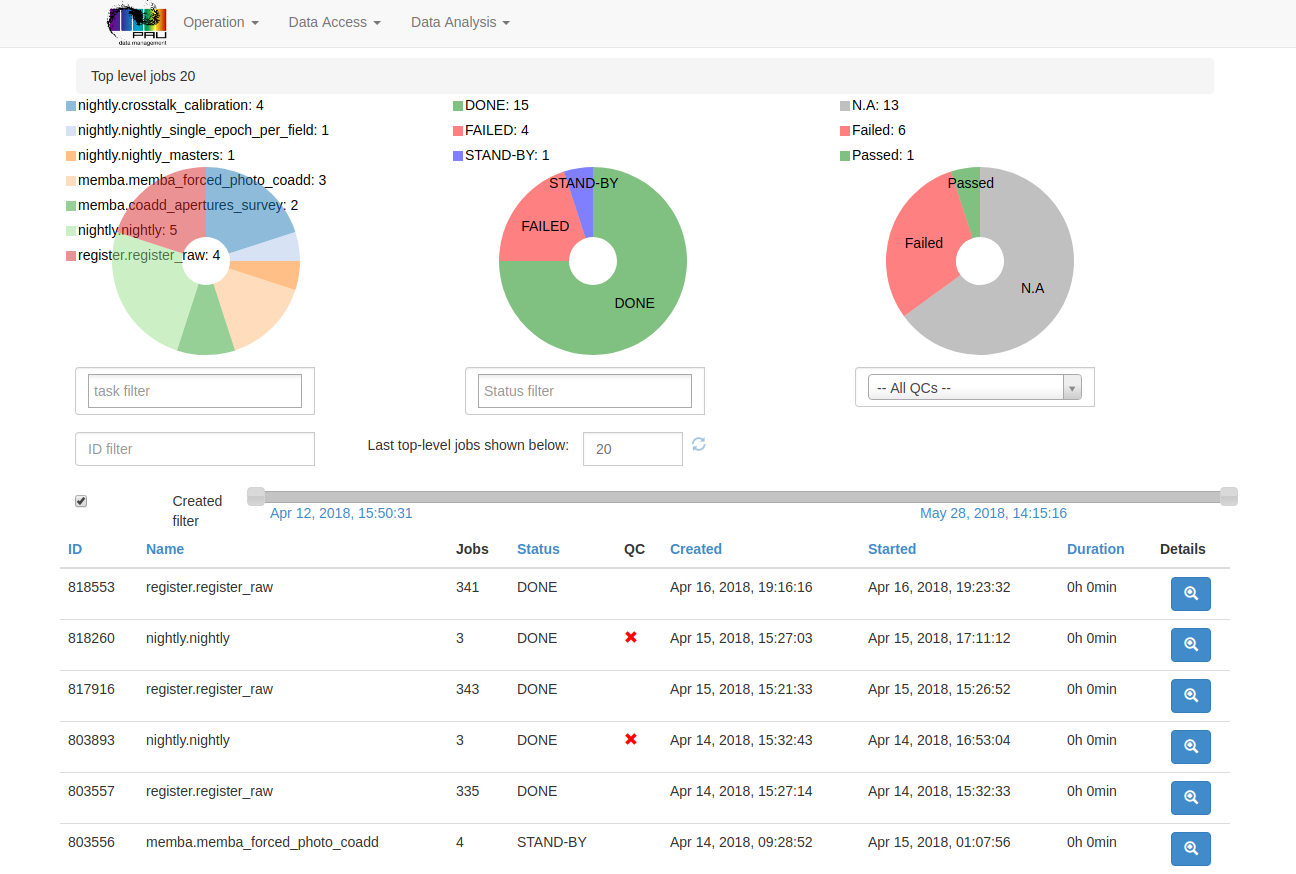}
\caption{View of the PAUS web interface for operations control. List of top-level tasks and their status.}\label{fig_operations1}
\end{figure*}

During operations, pilot jobs are launched in the PIC computing farm and the tasks are automatically fetched and executed. The execution status of the jobs can be monitored in the main section of the web interface, showing the results of the queue of batch jobs (see section \ref{s:orchestration}).

The list of the most recent (or current) top-level jobs is shown in table form listing ID, name, number of sub-jobs, execution status and quality control, as well as dates of
creation, start and duration (Figure \ref{fig_operations1}).
The list can be filtered by task, quality control, date range and status. Pie charts indicate the percentage of jobs for each task or status.

The details of each job that can be accessed from the web page are the configuration, input, output, log messages and traceback. In particular, the job configuration and its error traceback allow for a quick look in case of a failed job.

The jobs are hierarchically structured. Each job can have sub-jobs, with static dependencies, as a result of the parallelization of the pipeline execution. From the web, the job hierarchy is maintained and sub-jobs details can be accessed clicking on the top level jobs.

\begin{figure*}[!ht]
\centering
\vspace{0.8cm}
\includegraphics[width=1.5\columnwidth]{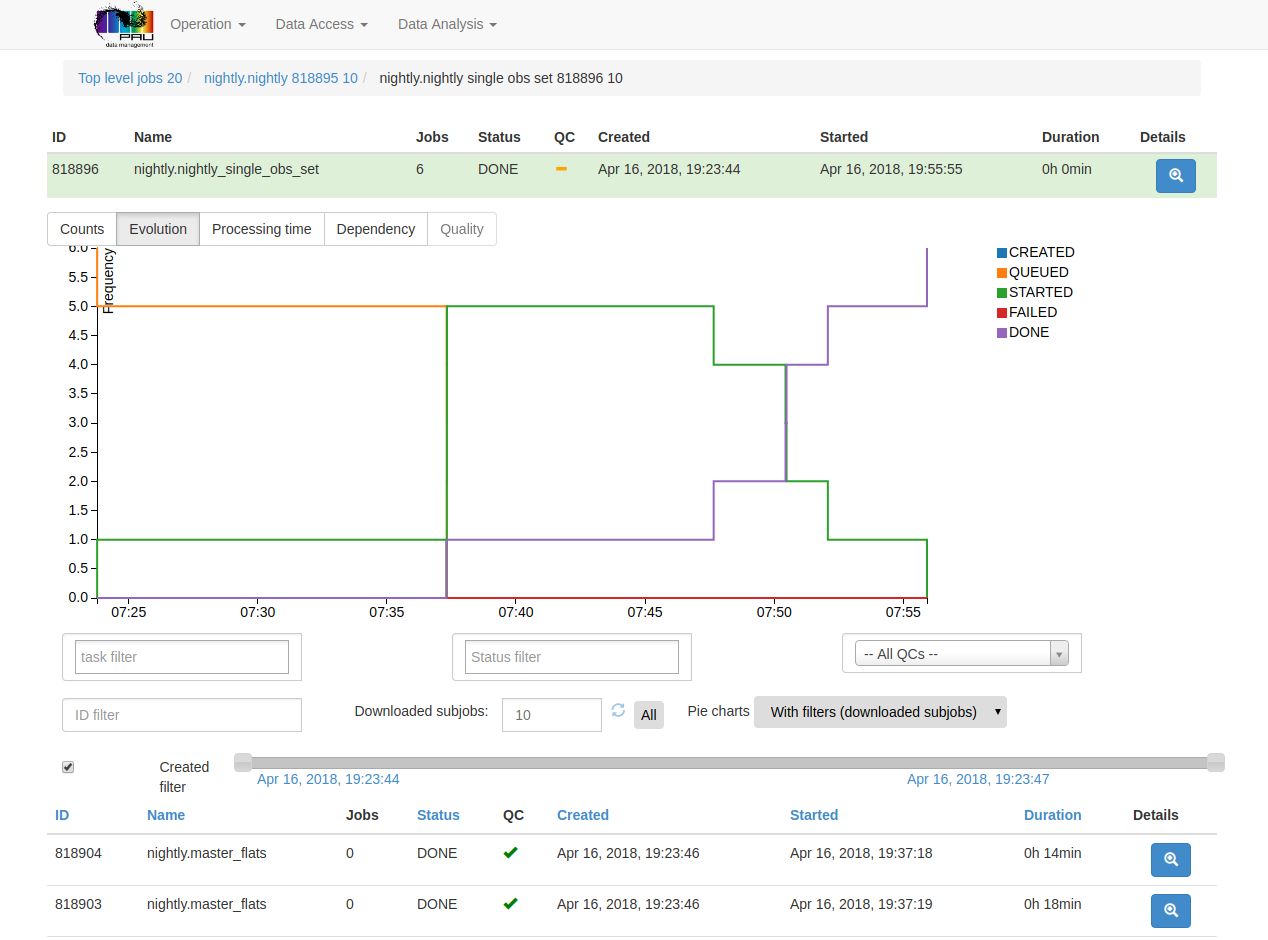}
\caption{View of the PAUS web interface for operations control. Time evolution of jobs processing one observation set.}\label{fig_operations2}
\end{figure*}

A series of graphical views are associated with the operation control web page. The execution status chart, for example, gives a quick overview of the time evolution of the jobs status of a certain pipeline (Figure \ref{fig_operations2}).

\subsection{Data Quality inspection}
\label{ss:data_quality}

During the execution of the tasks and in the epilog of the parent tasks, a series of quality checks are performed and its results assigned to the executed job. The quality checks can be either numeric values of parameters, to be compared to a certain value range, and/or plots for visual check. In case of parent jobs, quality controls plots are generated  for summarizing the evolution of interesting parameters calculated in their subtasks.

The results of the quality checks are registered in the PAU database and linked to the corresponding job, the plots are stored in a dedicated disk space accessible both from the nodes executing the jobs and from the web server. These results can be viewed through the job details page.

A table collects the name of each quality test, a short description of the check performed, the result value and the reference range value, a label that visually shows if the constraint is fulfilled or not, and finally a link to the corresponding plot (if available).

The quality inspection is especially interesting during the process of validation and debugging. During normal operations, the quality checks result is monitored, shown as a green, red or yellow (in case of partial, but not critical failure) flag in the main job operations page.

\subsection{Nightly Report}
\label{ss:nightly_report}

\begin{figure*}[!ht]
\centering
\includegraphics[width=1.5\columnwidth]{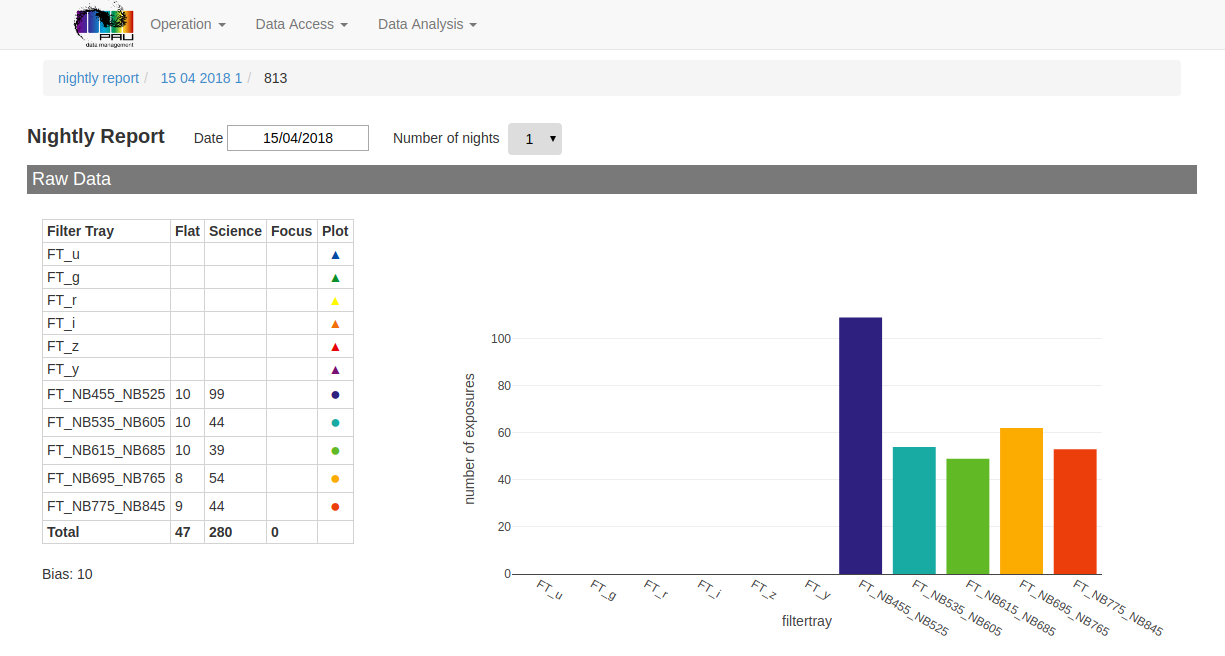}
\includegraphics[width=1.5\columnwidth]{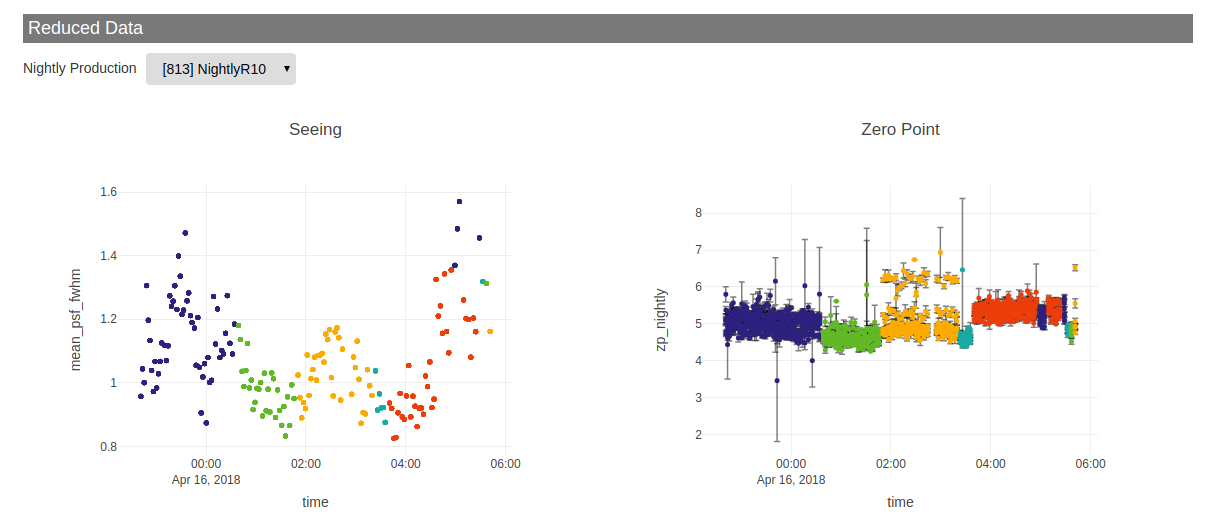}
\includegraphics[width=1.5\columnwidth]{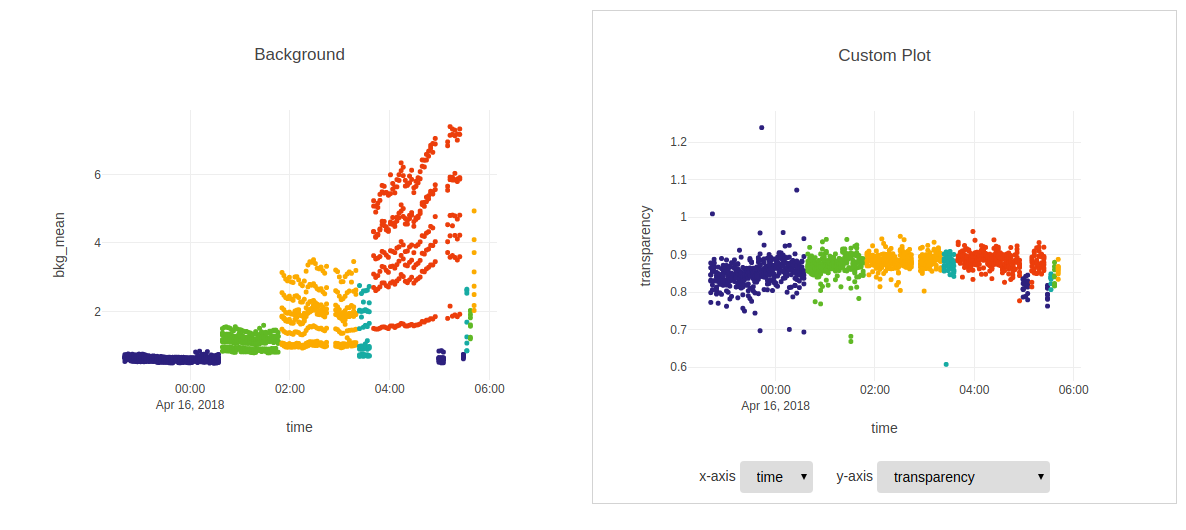}
\caption{PAUS web interface Nightly Report page example with plots.}\label{fig_nightlyreport1}
\end{figure*}

\begin{figure*}[!ht]
\centering
\includegraphics[width=1.5\columnwidth]{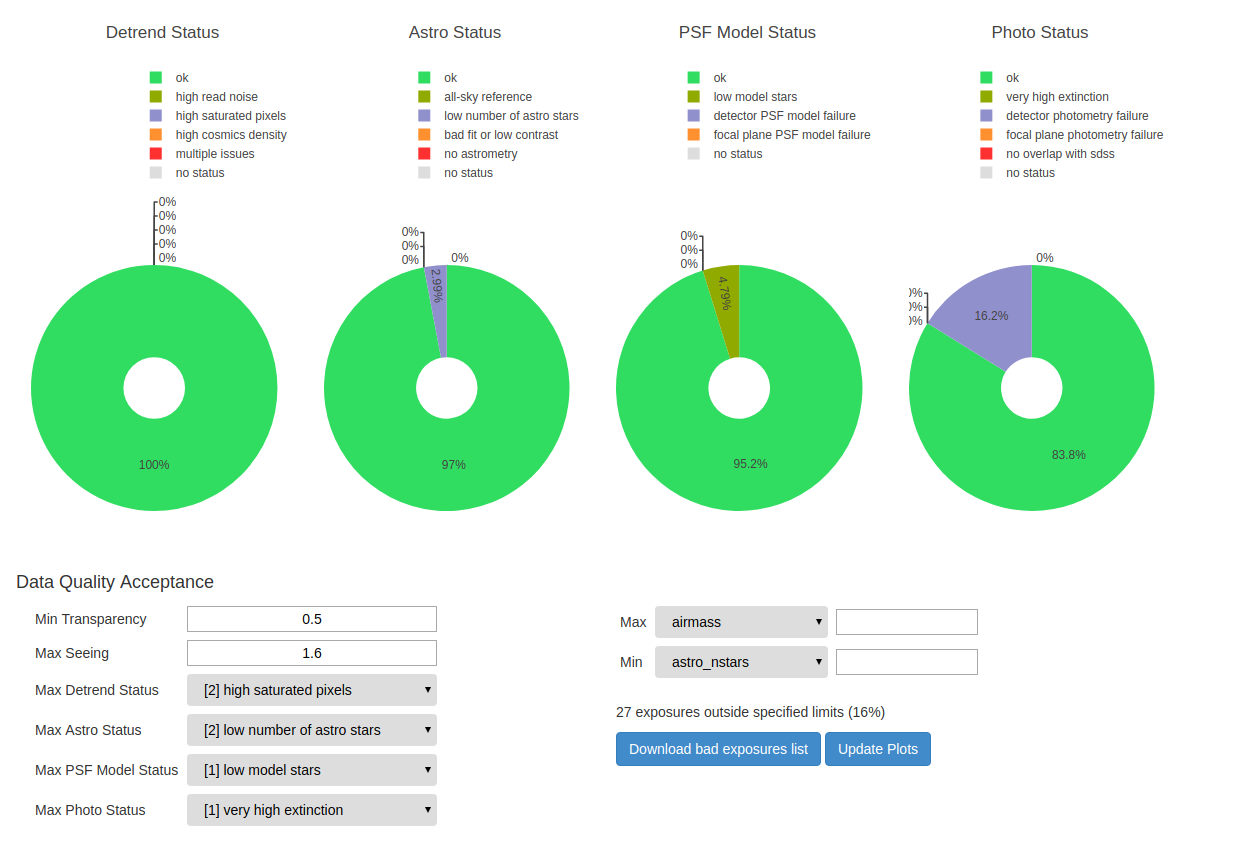}
\caption{PAUS web interface Nightly Report page example with quality selection of images.}\label{fig_nightlyreport2}
\end{figure*}

The main purpose of the Nightly Report web page is to feed into the survey and science program planning process.

It provides an overview of all the image parameters taken in a given range of observation nights and stored as raw metadata in PAUdb. An example is shown in Figure \ref{fig_nightlyreport1}. Additional metadata about seeing, sky background, detrending status etc. are available to be queried by the page after the nightly tasks have finished successfully and the information is safely stored in PAUdb.
These parameters are displayed as plots, showing the time evolution of observing conditions and data quality over the night.

Additional plots show the percentage of science frames that pass the quality control tests (Figure \ref{fig_nightlyreport2}), allowing the astronomer to assess the quality of the data taken during the previous observing night. Selection criteria are configurable in cut values and parameters, such as transparency, seeing, PSF, and parameters derived by the operations of detrending, astrometry and photometry. Those exposures that fail the quality control criteria can be filtered out and rescheduled updating the survey plan accordingly. 

\subsection{Data access and distribution}
\label{ss:data_access}

The PAUdb webpage allows the user to access the PAU files in archive, the PAUdb schema and the metadata through a SQL search window.

The PAU archive at PIC is accessible through WebDAV, a protocol offered by default by dCache, the PIC massive storage system. The access of the archive through WebDAV is especially intuitive in order to navigate the directories tree and download individual files (previously staged from tape to disk buffer). All the web page users have permission to navigate through the full archive, both raw and reduced data.

The production table is also published, with its full content. The Production index and/or the release name are the most common filters for searching data through all the different code versions that have been used to process the data. Each production is associated with a pipeline, and linked to the input pipeline through the field input\_production\_id.

The SQL search page connects to PAUdb with a read-only generic user. The construction of the query and the selection of the correct data production is supported by the publication of the full db schema: the name of all the tables, its fields, types, descriptions, and of the available indexes.

The SQL searches are limited in number of output rows, configurable by the user, but with an upper limit of 10k entries. The limitation is given by the performance and execution times of the platform on top of which the query runs. Anyway, PostgreSQL performance does not prevent having a very powerful tool to query the database, visualize the results on-line in form of table or plot (histogram, or scatter), create new fields applying functions or combine one or more fields, and download the result of the query in CSV format\footnote{\url{https://tools.ietf.org/html/rfc4180}}.
For wider searches it is possible to use CosmoHub, as explained in section \ref{s:cosmohub}.

\subsection{Aperture inspector and reporting system}
\label{aperture_inspector}

\begin{figure*}[!ht]
\centering
\includegraphics[width=1.25\columnwidth]{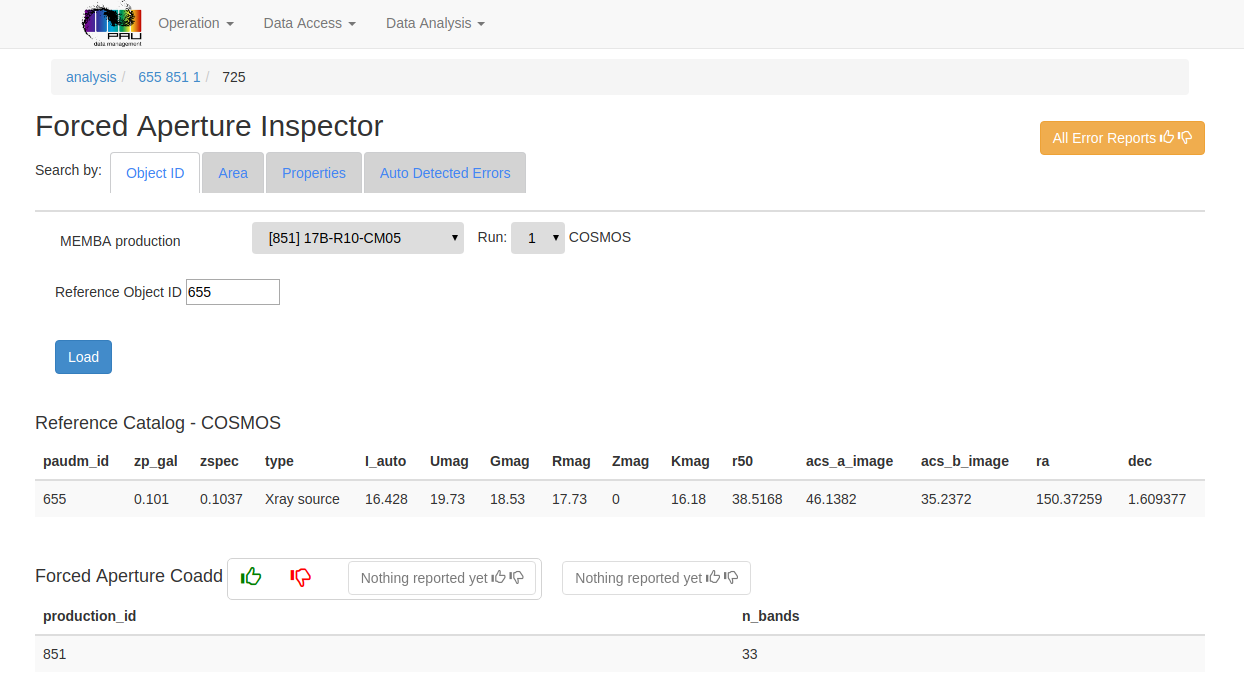}
\includegraphics[width=1.25\columnwidth]{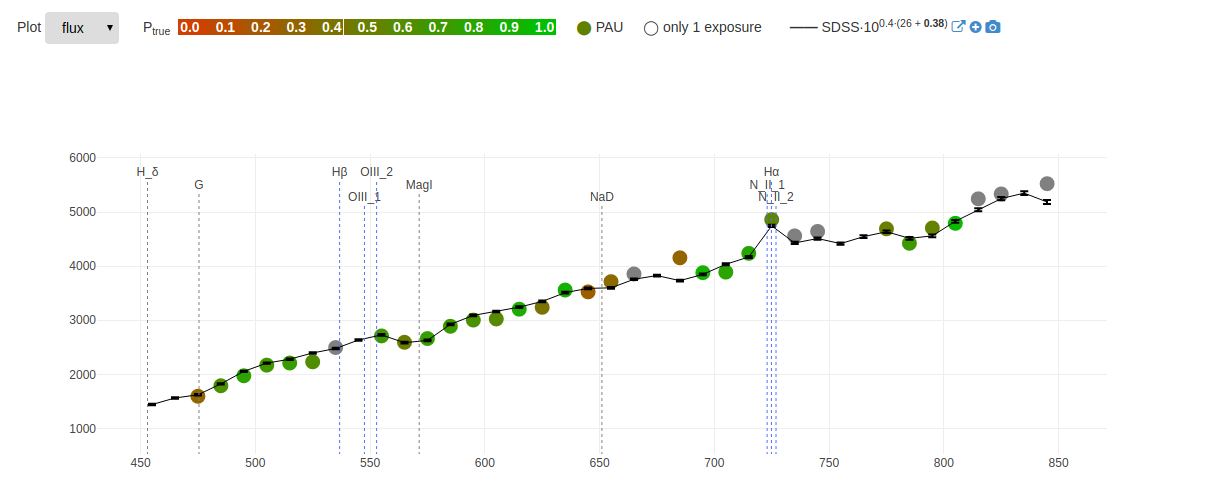}
\includegraphics[width=1.25\columnwidth]{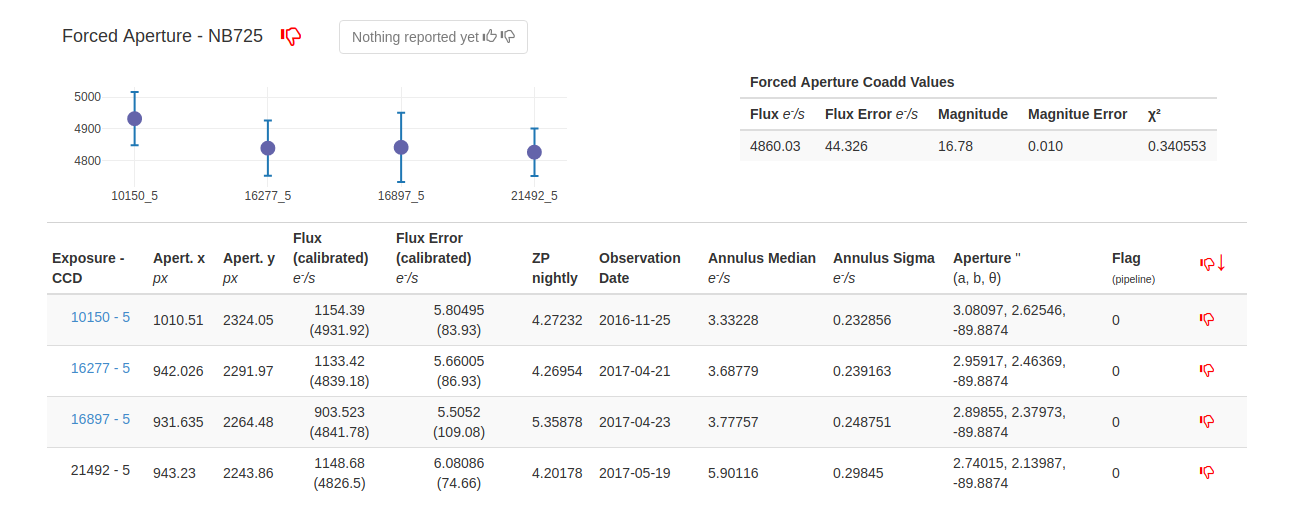}
\includegraphics[width=1.25\columnwidth]{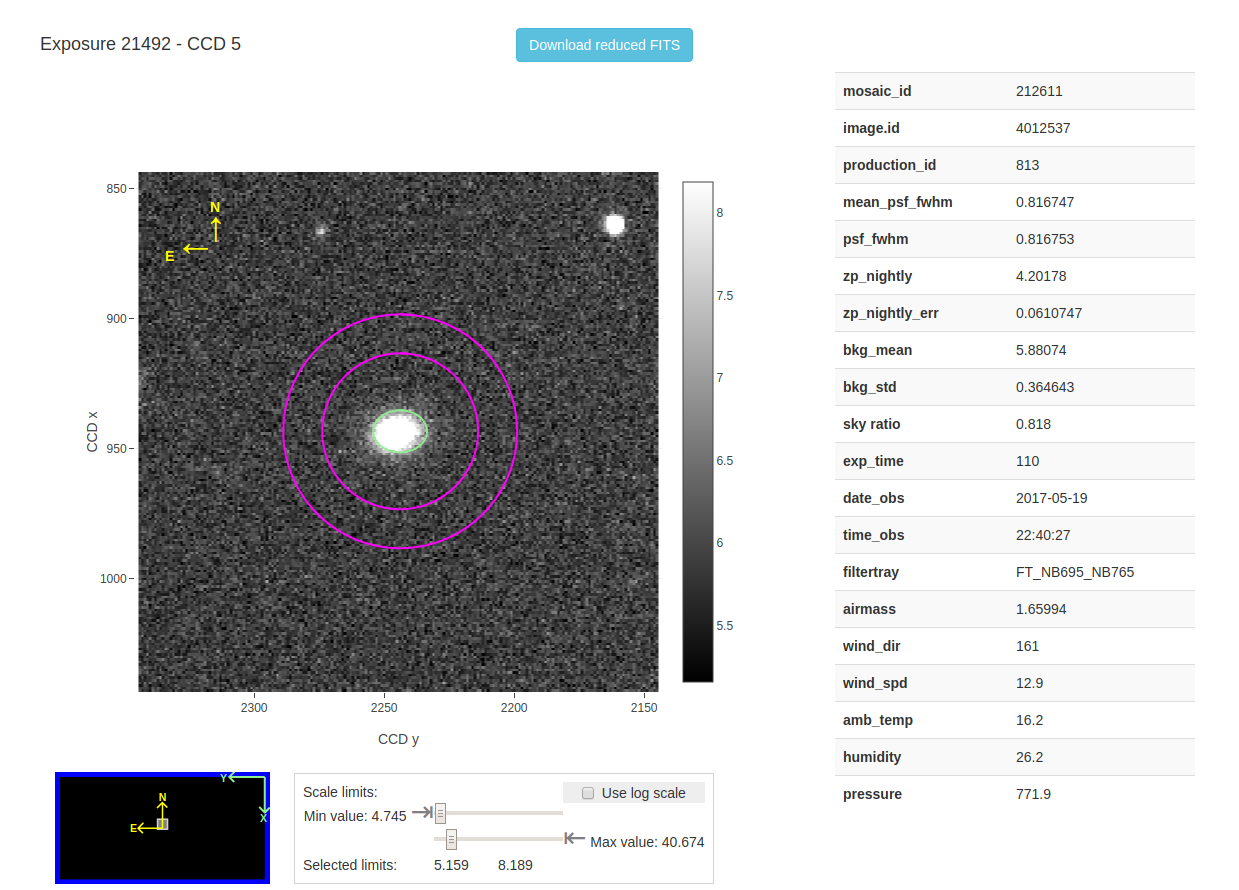}
\caption{PAUS web interface aperture inspector page example.}\label{fig_inspector}
\end{figure*}

The web interface is used to review the \textit{memba} pipeline output, i.e. the combined flux measurements obtained through the forced aperture photometry method for all bands and for each source measurement.

A dedicated page of the web site allows the user to relate the \textit{memba} result to all the pipeline reduction steps backwards down to any of the original exposures, in a visual way, providing plots of many of the individual measurements that contribute to the final result (for the details of the reduction and analysis steps see \citet{paudm1} and \citet{paudm2}). An example is shown in Figure \ref{fig_inspector}.

The user can select a source from the reference catalog, either by its reference ID, or search criteria such as area (e.g. RA, Dec), parameters of the source (e.g. magnitude in a given band) or the PAUdm reduction results (e.g. flux limit in a given narrow band), etc.

Once an object is selected and loaded its 40 co-added narrow band fluxes are displayed as in a wavelength-flux plot.
If an SDSS spectrum is available\footnote{See \url{https://www.sdss.org/dr14/spectro/} for the methods to access the published SDSS spectroscopic data (DR14).} for that object, 40 corresponding SDSS measurements, derived by convolving the SDSS spectrum with the filter response functions of the 40 PAU narrow band filters, are superimposed alongside with emission and absorption lines from a line catalog.

This provides an initial check of how well PAUdm results match results from spectroscopy surveys, and how well it picks up spectral lines.

Each of the narrow band measurements is selectable. On selection, a list of all forced aperture measurements that contributed to the overall co-added value is loaded.
The user can inspect parameters specific to these individual measurements, e.g. flux, aperture size
and orientation, sky background and observation date etc.

For each individual measure it is possible to visualize an image section around the source with the superimposition of the aperture region used to calculate the object flux and the annulus selected to calculate the sky background. Metadata related to the weather conditions and telescope parameters during the observations are loaded and displayed.

\subsection{Catalogs distribution}
\label{s:cosmohub}

\begin{figure*}[!ht]
\centering
\includegraphics[width=1.9\columnwidth]{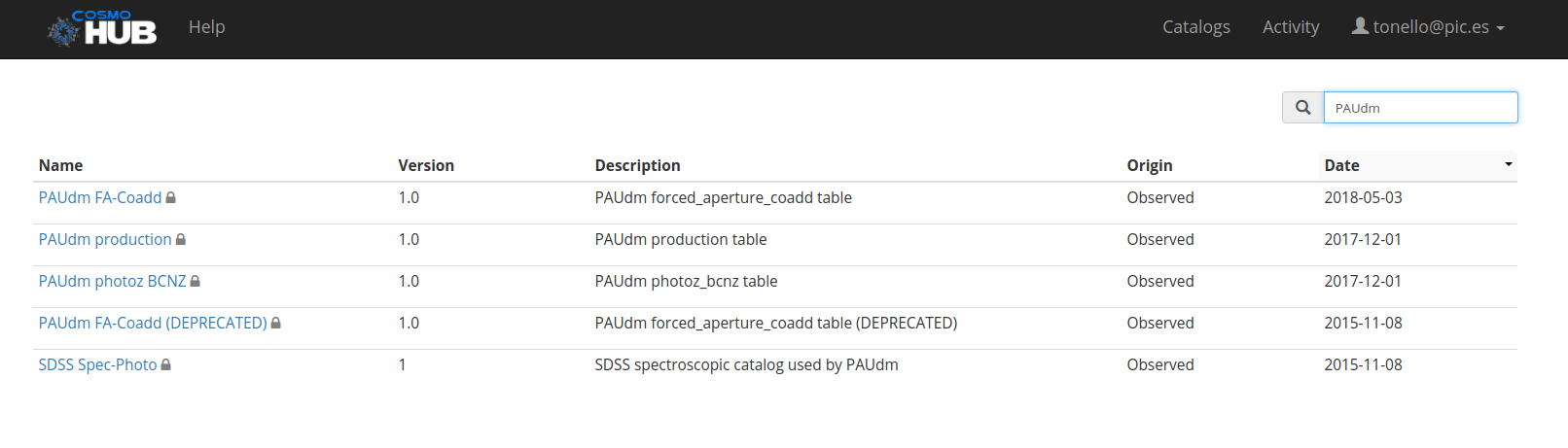}
\caption{PAUdm catalogs distributed in CosmoHub.}\label{fig_chub1}
\end{figure*}

\begin{figure*}[!ht]
\centering
\includegraphics[width=1.4\columnwidth]{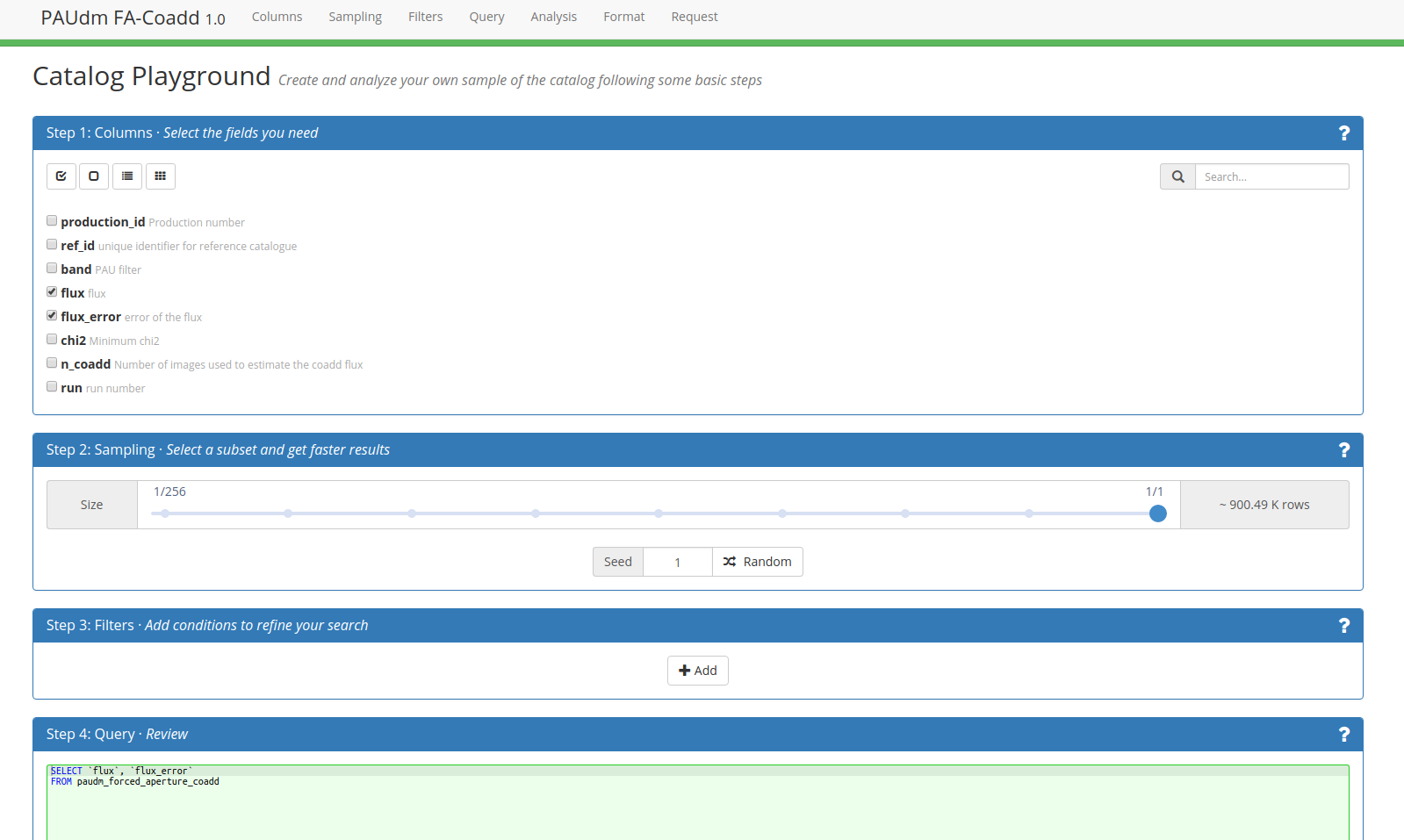}
\includegraphics[width=1.4\columnwidth]{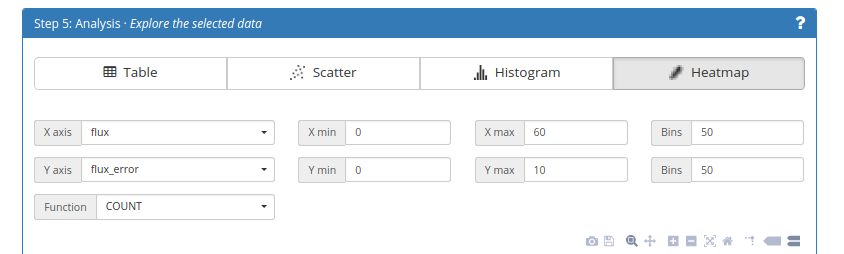}
\includegraphics[width=1.385\columnwidth]{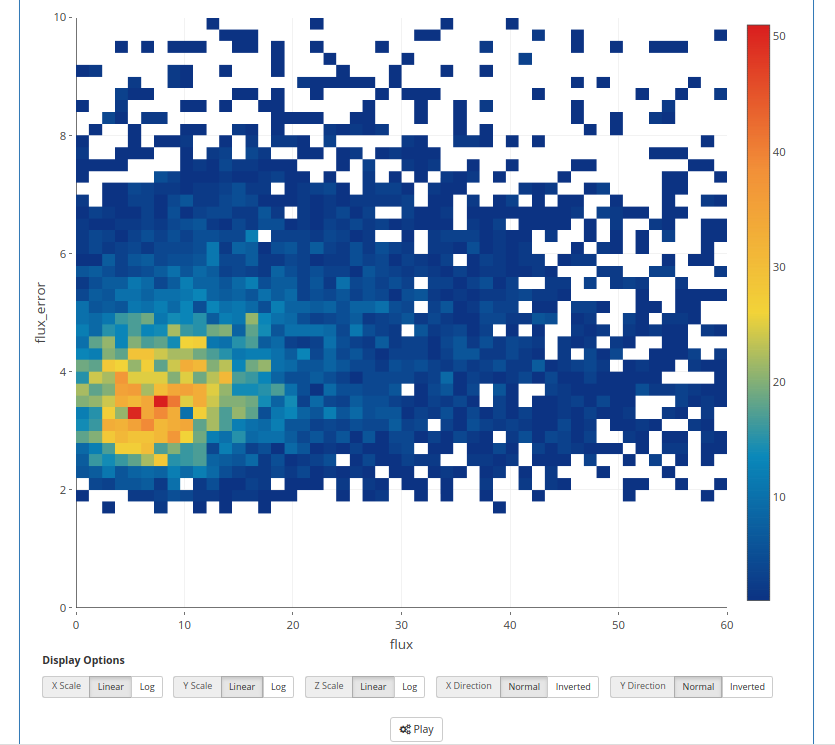}
\caption{CosmoHub for PAUdm catalogs distribution and inspection example.}\label{fig_chub2}
\end{figure*}

One of the tools we have adopted to access and distribute data for the PAUS collaboration is CosmoHub\footnote{\url{https://cosmohub.pic.es/}}). 

CosmoHub is a web platform based on big data technologies developed at PIC to perform interactive exploration and distribution of massive cosmological datasets without any SQL knowledge being required. The latest release has been built on top of Apache Hive, a data warehouse based on Apache Hadoop which facilitates reading, writing and managing large datasets.

CosmoHub is hosted at the Port de Informaci\'{o} Cient\'{i}fica (PIC) and provides support not only to PAU Survey, but also to several international cosmology projects such as the ESA \textit{Euclid} space mission, the Dark Energy Survey (DES) and the Marenostrum Institut de Ci\`{e}ncies de l'Espai Simulations (MICE).

CosmoHub allows users to access value-added data, which usually are complementary files to analyze the data such as sky or survey properties maps, to load and explore pre-built datasets and to create customized object catalogs through a guided process. All those datasets can be interactively explored using an integrated visualization tool which includes 1D histogram and 2D heatmap plots (Figure \ref{fig_chub2}). Finally, all those datasets can be downloaded in standard formats.

We currently ingest into CosmoHub three tables of PAUdb (see a screenshot in Figure \ref{fig_chub1}), containing metadata of memba pipeline productions (table production and forced\_aperture\_coadd) and photometric redshift (photo-z) results. Data from external surveys, used to calibrate, to perform forced photometry or to cross-check PAUS data (such as COSMOS, CFHTLenS or SDSS) is also accessible from CosmoHub for a direct comparison, in addition to mock galaxy catalogs created by the MICE collaboration, which are used for calibrating and testing the different pipelines. The access to this information allows the PAU Survey collaborators to explore and download the PAUdm pipelines output data for the scientific exploitation.

%--------------------------
\section{PAUS data center}
\label{s:pic}

\begin{figure*}[!ht]
\centering
\includegraphics[width=2\columnwidth]{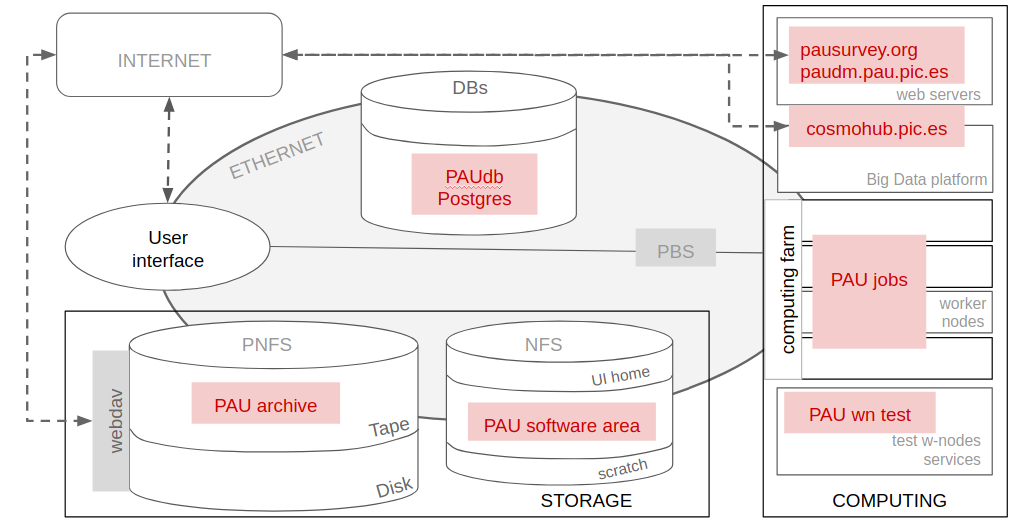}
\caption{PAU infrastructure and services. For a detailed description of the PIC infrastructure see appendix A.}\label{fig_pic}
\end{figure*}

PAU data management has been developed on top of the PIC infrastructure, which consists of a storage system, a network, a computing farm, databases and user interfaces (UI). Details of the PIC infrastructure are described in Appendix A.

The PAU project has access to a series of standard and customized services (user interface, test worker node, web servers on virtual machines and databases), supporting the activities of the data management. The standard user interface system of PIC allows any user of the PAU project to access the PIC infrastructure, with a dedicated 10 GiB "home" space, a shared scratch NFS area, and access to the PAU files archive. An interactive access to a test worker node offers an environment which allows developers to test and debug the PAU pipelines code before the production phase. 

For computing operations, the PIC farm is shared among the hosted projects. 5\% of the 8000 available cores at PIC (year 2018) have been assigned to PAUdm activities. The software developed for PAU Survey image reduction and analysis is managed through git, accessible from an NFS software area, readable by all the nodes where jobs run.

Raw data from the WHT is transferred to PIC through a 10 Gb connection (further details on the PIC network connections can be found in Appendix A).

PAU files are permanently stored on tape. An automated double copy, in two different cartridges, guarantees the preservation of PAU Survey data in case of failure of one of them. The PAUCam data is kept in a disk buffer for a short period of time after their arrival at PIC, in order to read and process it as soon as it is transferred successfully.

The virtualized infrastructure of PIC hosts the web services for PAU Survey: the official PAU Survey web site, and the internal web service\footnote{\url{http://paudm.pau.pic.es}}, entirely dedicated to PAUdm activities (see section \ref{s:web_services}). 

The PIC storage manager dCache integrates WebDAV, among others, as a native protocol for data access. In addition to its use for archive files access from the internal PAUdm web pages, the WebDAV protocol allows the groups that do not belong to PAU Survey, and that received a time allocation to use PAUCam for their own scientific purposes, restricted access to download the raw and reduced images belonging to their program.

Once PAU Survey raw and reduced data become public, it will be possible to have open access to it through WebDAV. 

%-----------------------------------
\section{Conclusions}
\label{conclusions}

The PAUS data management team activity is one of the key points for the success of the PAU Survey project. It is in charge of the data transfer from the observatory to the data center, the deployment and the execution of the \textit{nightly} pipeline and other analysis pipelines, as well as the organization and distribution of the final results and the metadata produced during the data process.

The PAUS PostgreSQL relational database is the solution adopted by the project for the PAUdm operation and orchestration. It has been designed to be the pillar of the data management, not only to preserve information but also its consistency, from the raw metadata to the final catalogs and results reproducibility, keeping track of the archive status, the processes run, the metadata and final catalogs. It has been developed with the idea of helping the data accessibility, as well as the distribution of the final catalogs.

PostgreSQL has been proven to be a good choice for frequent data insertion, delete, update, but with limits when having more than hundreds concurrent connections, blocking operations or updating status of thousands of subjobs. \citet{platform} shows PostgreSQL limits when querying data volumes larger than some hundreds of GiB.

CosmoHub has been a natural solution for catalogs exploration and distribution, but the process of PAUdb tables migration to this platform is still manual. We are exploring the way of an automatic insertion of the \textit{memba} PAUdm pipeline output to CosmoHub. Synchronizing contents between the PostgreSQL PAUS database and CosmoHub will not only improve the handling of PAUS catalogs, but also their findability, once the standardization of CosmoHub content in the Virtual Observatoy (VO) paradigm will be completed.

The orchestration of the PAU data management tasks through the novel developed tool BT has been widely and successfully used for the nightly operations and jobs orchestration of PAUdm in the HTC infrastructure at PIC. The size of the project, the cost of design, deploying, maintenance and upgrade a software tool like BT, and the availability of generic middleware tools for managing jobs (for example HTCondor\footnote{\url{https://research.cs.wisc.edu/htcondor/}}) was discouraging ad-hoc self-designed solutions. The advantage of a tool like BT is that, although built ad-hoc for the PAUdm pipelines case, it has been designed, developed and used in a widely and more generic context, proving its versatility and effectiveness in the interaction of complex tasks structures with the HTC environment. 

The performance of the automatic transfer procedure has been demonstrated to fulfill the PAUS project needs. The download of the data to analyze from the observatory in La Palma to PIC finalizes in a few hours, with total reliability during normal operations, and fully recovers in the rare cases of failure. While more sophisticated tools would allow a total reliability and automation, the data volume to be transferred per night and the lifetime of the project lead us to go for a simple and easily manageable solution.

The PIC storage system for PAU images has been proved to be reliable, with total accessibility of the files stored in magnetic tapes and recover capability. The low access rate of PAUS images by the \textit{nightly} and \textit{memba} pipelines justify the use of this cheap, though slow, data storage support, with respect to other supports like hard disks or solid state drives.

The PIC computer farm has been able to process PAU jobs according to the expectations. The high level of parallelization of the pipelines (see Figure \ref{fig_nightlydataflow}) in independent jobs, the number of computing nodes and the distributed file system available at PIC for the PAUdm jobs process allows us to fulfill the requirement of having nightly results and data quality available on time for optimizing the survey program. The HTC infrastructure enables the reprocessing of all the observations of the Survey in few days, confirming it to be a valid choice.

Web services have been implemented for PAU data and metadata access, retrieval and analysis. They have received a very positive feedback from the collaborators, due to their usability and information content. The data services implemented are giving a valid and reliable support to the first scientific production exploiting the PAUS data (see \citet{lee, luca, laura}).

\section*{Acknowledgements}

Funding for PAUS has been provided by Durham University (via the ERC StG DEGAS-259586), ETH Zurich, Leiden University (via ERC StG ADULT-279396 and Netherlands Organisation for Scientific Research (NWO) Vici grant 639.043.512) and University College London. The PAUS participants from Spanish institutions are partially supported by MINECO under grants CSD2007-00060, AYA2015-71825, ESP2015-66861, FPA2015-68048, SEV-2016-0588, SEV-2016-0597, and MDM-2015-0509, some of which include ERDF funds from the European Union. IEEC and IFAE are partially funded by the CERCA program of the Generalitat de Catalunya. The PAU data center is hosted by the Port d'Informaci\'o Cient\'ifica (PIC), maintained through a collaboration of CIEMAT and IFAE, with additional support from Universitat Aut\`onoma de Barcelona and ERDF. We aknowledge the PIC services department team for the support work and the fruitful discussions.



%% References


\section*{References}

\bibliography{sample.bib}


%%\section*{References}

%%\bibitem{paudm1}
%%PAU collaboration, "PAUdm calibration", in preparation
% \bibitem{paudm2}
% PAU collaboration, "PAUdm photometry ", in preparation

% \end{thebibliography}

\appendix
\section*{Appendix A: PIC infrastructure}
\label{appendix:pic}

The Port d'Informaci\'o Cient\'ifica (PIC) is a research support center that offers services to manage large amounts of data to scientific collaborations, whose researchers are spread all over the world, using distributed computing technologies including clusters, grid, cloud and big data.

PIC has two different vaults in the same building, located in the campus of the Universitat Aut\'onoma de Barcelona, with different characteristics and energy efficiency profiles: 
a 150 m\textsuperscript{2} air-cooled room and 
a 25 m\textsuperscript{2} highly energy efficient\citet{acin} room which uses open bath dielectric fluid tanks for the storage and computing IT equipment.

The PIC infrastructure offers a hierarchical mass storage service with a capacity of approximately 7 PiB on disk and 18 PiB on magnetic tape. The disk pools are managed by dCache/Chimera\footnote{https://www.dcache.org/}. Shared scratch disk space is also available (Network File System, NFS) to support operations.

Long term storage uses tapes, organized in a robotic tape library and managed by Enstore\footnote{http://www-ccf.fnal.gov/enstore/}. The magnetic tape technology is constantly kept up to date. The technology currently in use (year 2018) is the Oracle T10000T2\footnote{http://www.oracle.com/us/products/servers-storage/storage/tape-storage/storagetek-t10000-t2-cartridge-296699.pdf} with a capacity of 8.5 TiB (non compressed) per cartridge. The reliability of the magnetic support is extremely high (warranty of uncorrected bit error rate $1\times10^{-19}$, 30 years of archival life and 25~000 loads/unloads). Data in tape is pre-staged in a disk buffer for fast access, totally transparent for the user.

The PIC computing batch system is based on PBS Torque/MAUI for the job queue system and scheduling, and HTCondor. It is integrated to the WLCG (WorldWide LHC Computing Grid), and it is also accessible by local users. The PIC's computing cluster consists of 8000 cores running on Linux servers. 

The external network is deployed in collaboration between the Catalan NREN (CSUC, Anella Científica), the Spanish NREN (RedIRIS) and the Géant pan-European network. Being the Spanish Tier-1 center for CERN, PIC is connected through a 20 Gb link between PIC, CERN and the other Tier-1s. It is also connected to the LHCONE network to Tier-2s and to La Palma, one of the Canary Islands where the Roque de los Muchachos Observatory is located.

PIC offers a series of services (user interface, test worker node, virtual machines), supporting the activities of the data management of the scientific projects, mainly of particle physics (Atlas, CMS. LHCb), astrophysics (MAGIC, CTA) and cosmology (PAU Survey, \textit{Euclid}, MICE).

The user interface system of PIC allows any user to access the PIC infrastructure, with a dedicated "home" space, access to a shared scratch NFS area, access to the storage system and to customized front-ends in order to optimize the delivery of users’ scientific results.
The virtualized infrastructure of PIC is based on Ovirt.

PIC infrastructure includes a big data platform based on Hadoop (Hortonworks HDP v2.6), deployed at PIC for simulated and observed cosmological catalogs analysis, with a dedicated web portal called CosmoHub (see section \ref{s:cosmohub}) for data inspection and distribution.

PIC is involved in the EU HNSciCloud Pre-Commercial Procurement project lead by CERN in order to deploy hybrid cloud prototypes oriented to science needs.

%%\end{linenumbers}

\section*{Appendix B: PAU database schema}
\label{appendix:paudb}

In the following we show the names, description and the columns list of each of the main tables of the PAUdb schema. Foreign keys, which determine the relation between tables, are marked in \textit{italic}.

\begin{table*}
\caption{PAUdb tables}
\label{tab:paudm}
\begin{tabular}{| p{.3\columnwidth} | p{.5\columnwidth} |p{1.1\columnwidth}| } 
\hline
\textbf{Table name} & \textbf{Description} & \textbf{Columns} \\
\hline
\textbf{production}  & Tracks the different processing production runs for all pipelines &  comments, created, id, \textit{input\_production\_id}, \textit{job\_id}, pipeline, release, software\_version \\
\hline
\textbf{mosaic}  & List of mosaic exposure images (raw and reduced) &  airmass, amb\_temp, archivepath, astro\_chi2, astro\_contrast, astro\_href\_sigma, astro\_nstars, astro\_nstars\_highsn, astro\_ref\_cat, astro\_ref\_sigma, astro\_status, comment, date\_creat, date\_obs, dec, detrend\_status, equinox, exp\_num, exp\_time, extinction, extinction\_err, filename, filtertray, filtertray\_tmp, guide\_enabled, guide\_fwhm, guide\_var, humidity, id, instrument, kind, mean\_psf\_fwhm, merged\_mosaics, nextend, \textit{obs\_set\_id}, obs\_title, photo\_status, pressure, \textit{production\_id}, psf\_model\_status, ra, rjd\_obs, telfocus, time\_creat, time\_obs, wind\_dir, wind\_spd, \textit{zp\_phot\_id} \\ 
\hline
\textbf{image} & List of images associated with the mosaics (CCD and single amplifier images) &  amp\_num, bkg\_mean , bkg\_std, ccd\_num, cosmic\_ratio, dec\_max, dec\_min, filter, gain, id, image\_num, max\_readnoise, \textit{mosaic\_id}, naxis1, naxis2, n\_extracted, psf\_fit, psf\_fwhm, psf\_stars, ra\_max, ra\_min, rdnoise, saturate\_ratio, transparency, waveband, wavelength, zp\_nightly, zp\_nightly\_err, zp\_nightly\_stars \\
\hline
\textbf{obs\_set}  & List of Observation Sets registered in the database &  id, instrument, log, night, notes, obs\_set, operator, rjd\_start, rjd\_stop \\
\hline
\textbf{project} & Description of projects observing with PAUCam & contact\_email, contact\_name, created\_at, description, id, name \\
\hline
\textbf{crosstalk\_ratio}  & Crosstalk correction to be applied to each amplifier &  amp\_num\_dest, amp\_num\_orig, ccd\_num\_dest, ccd\_num\_orig, \textit{production\_id}, ratio \\
\hline
\textbf{detection} & Detections measured directly on the image after the nightly data reduction & id, \textbf{image\_id}, insert\_date, band, background, class\_star, spread\_model, spreaderr\_model, flux\_auto, flux\_err\_auto, flux\_psf, flux\_err\_psf, flux\_model, flux\_err\_model, flags, elongation, dec, ra, x, y, zp\_offset \\
\hline
\end{tabular}
\end{table*}

\begin{table*}
\caption{PAUdb nightly calibration tables}
\label{tab:paudm_cal}
\begin{tabular}{| p{.3\columnwidth} | p{.5\columnwidth} |p{1.1\columnwidth}| } 
\hline
\textbf{Table name} & \textbf{Description} & \textbf{Columns} \\
\hline
\textbf{phot\_method}  & Photometric methods & background\_method, background\_parameter, comments, extraction\_code, extraction\_method, extraction\_parameter, id, scatterlight\_method, scatterlight\_parameter \\
\hline
\textbf{phot\_zp}  & Photometric zero-points & band, date, id, \textit{production\_id}, zp \\
\hline
\textbf{template} & SED of star templates references & filename, id, template\_index, template\_lib, template\_name \\
\hline
\textbf{star\_photometry} & Calibration stars fluxes & bg, bg\_err, flags, flux, flux\_err, id, \textit{image\_id}, \textit{phot\_method\_id}, ref\_cat, ref\_id, x\_image, y\_image \\
\hline
\textbf{star\_template\_zp} & SED template associated with each star &  id, \textit{star\_zp\_id}, \textit{template\_fit\_band\_id}, zp, zp\_error, zp\_weight \\
\hline
\textbf{star\_zp}  & Star zero points calculated in the nightly calibration &  calib\_method, id, \textit{star\_photometry\_id}, zp, zp\_error, zp\_weight \\
\hline
\textbf{image\_zp} & Image zeropoint measurements for each photometry-calibration method &  \textit{calib\_method, id}, \textit{image\_id}, \textit{phot\_method\_id}, transparency, zp, zp\_error \\
\hline
\end{tabular}
\end{table*}

\begin{table*}
\caption{PAUdb \textit{memba}  and catalogs tables}
\label{tab:memba}
\begin{tabular}{| p{.30\columnwidth} | p{.5\columnwidth} |p{1.1\columnwidth}| } 
\hline
\textbf{Table name} & \textbf{Description} & \textbf{Columns} \\
\hline
\textbf{forced\_aperture}  & Contains the single measurements using force photometry in \textit{memba} for each band and pass, and for each reference source &  annulus\_a\_in, annulus\_a\_out, annulus\_b\_in, annulus\_b\_out, annulus\_ellipticity, annulus\_median, annulus\_samples, annulus\_sigma,
aperture\_a, aperture\_b, aperture\_theta, aperture\_x, aperture\_y, flag, flux, flux\_error, image\_ellipticity, \textit{image\_id}, pixel\_id,
\textit{production\_id}, ref\_id \\
\hline
\textbf{forced\_aperture\_coadd}  & Contains the combined measurements using force photometry in \textit{memba} for each band and for each reference source & band, chi2, flux, flux\_error, n\_coadd, \textit{production\_id}, ref\_id, run \\
\hline
\textbf{sdss\_spec\_photo} & & \_class, dec, extinction\_g, extinction\_i, extinction\_r, extinction\_u, extinction\_z, fiberID, fiberMagErr\_g, fiberMagErr\_i, fiberMagErr\_r, fiberMagErr\_u, fiberMagErr\_z, fiberMag\_g, fiberMag\_i, fiberMag\_r, fiberMag\_u, fiberMag\_z, mjd, mode, modelMagErr\_g, modelMagErr\_i, modelMagErr\_r, modelMagErr\_u, modelMagErr\_z, modelMag\_g, modelMag\_i, modelMag\_r, modelMag\_u, modelMag\_z, objID, plate, ra, specObjID, subClass, survey, tile, z, zErr, zWarning \\
\hline
\textbf{memba\_ref\_cat}  & Reference catalog used for a given memba production &  \textit{production\_id}, ref\_cat \\
\hline
\textbf{photoz\_bcnz}  & Photometric redshifts &  chi2, ebv, n\_band, odds, \textit{production\_id}, pz\_width, ref\_id, zb, zb\_mean \\
\hline

\end{tabular}
\end{table*}

\begin{table*}
\caption{PAUdb BT tables}
\label{tab:bt}
\begin{tabular}{| p{.3\columnwidth} | p{.5\columnwidth} |p{1.1\columnwidth} |} 
\hline
\textbf{Table name} & \textbf{Description} & \textbf{Columns} \\
\hline
\textbf{dependency}  &  Tracks the dependency between Brownthrower jobs (Operation table) & \textit{super\_id}, \textit{parent\_id}, \textit{child\_id}\\
\hline
\textbf{job}  & Tracks the list of Brownthrower computing jobs (Operation table) &  config, description, id, input, name, output, status, \textit{super\_id}, token, ts\_created, ts\_ended, ts\_queued, ts\_started \\
\hline
\textbf{quality\_control} & quality control entries measured during the data reduction process & check\_name, id, \textit{job\_id}, max\_value, min\_value, plot\_file, qc\_pass, ref, time, units, value \\
\hline
\textbf{tag} & Configurable tags associated with a job (tracebacks, logs, etc.) &  \textit{job\_id}, name, value \\
\hline
\end{tabular}
\end{table*}

\begin{table*}
  \caption{PAUdb Public external tables}
  \label{tab:public}
 \begin{tabular}
  {| p{.30\columnwidth} | p{.5\columnwidth} |p{1.1\columnwidth} |} 
  \hline
\textbf{Table name} & \textbf{Description} & \textbf{Columns} \\
\hline
\textbf{cosmos} & External table from zCOSMOS. Sources with accurate redshifts for forced photometry and validation & acs\_a\_image, acs\_b\_image, acs\_mag\_auto, acs\_magerr\_auto, acs\_theta\_image, Bmag, conf, dchi, dec, ebv\_gal, ebv\_int, eI, F814W, Gmag, I\_auto, ICmag, Imag, Jmag, Kmag, mod\_gal, MV, NbFilt, paudm\_id, r50, ra, Rmag, sersic\_n\_gim2d, type, Umag, Vmag, zfits, zl68\_gal, zl99\_gal, Zmag, zp\_gal, zp\_sec, zspec, zu68\_gal, zu99\_gal \\
\hline

\textbf{sdss\_spec\_photo}  & External table from SDSS DR12 (SpecPhoto view). Sources with spectra for forced photometry and validation &  \_class, dec, extinction\_g, extinction\_i, extinction\_r, extinction\_u, extinction\_z, fiberID, fiberMagErr\_g, fiberMagErr\_i, fiberMagErr\_r, fiberMagErr\_u, fiberMagErr\_z, fiberMag\_g, fiberMag\_i, fiberMag\_r, fiberMag\_u, fiberMag\_z, mjd, mode, modelMagErr\_g, modelMagErr\_i, modelMagErr\_r, modelMagErr\_u, modelMagErr\_z, modelMag\_g, modelMag\_i, modelMag\_r, modelMag\_u, modelMag\_z, objID, plate, ra, specObjID, subClass, survey, tile, z, zErr, zWarning \\
\hline
\textbf{spec\_conv}  & Convolved fluxes derived from spectra observations from external surveys (i.e. SDSS, COSMOS, DEEP2, etc. &  band, flux, flux\_err, id, instrument, spec\_cat, spec\_id \\
\hline
\textbf{usnostars} &Stars from USNO catalog & dec, field, id, mag\_g, mag\_r, ra \\
\hline
\textbf{yalestars} & Stars from Yales catalog & dec, id, mag\_v, ra \\
\hline
\textbf{cfhtlens} & CFHTLenS catalogue for Forced Photometry & alpha\_j2000, a\_world, backgr, bpz\_filt, bpz\_flagfilt, bpz\_nondetfilt, bulge-fraction, b\_world, c2, chi\_squared\_bpz, class\_star, delta\_j2000, e1, e2, erra\_world, errb\_world, errtheta\_j2000, extinction, extinction\_g, extinction\_i, extinction\_r, extinction\_u, extinction\_y, extinction\_z, field, field\_pos, fitclass, fit-probability, flag, flux\_radius, fwhm\_image, fwhm\_world, imaflags\_iso, imaflags\_iso\_g, imaflags\_iso\_i, imaflags\_iso\_r, imaflags\_iso\_u, imaflags\_iso\_y, imaflags\_iso\_z, isoarea\_world, kron\_radius, level, lp\_log10\_sm\_inf, lp\_log10\_sm\_med, lp\_log10\_sm\_sup, lp\_mg, lp\_mi, lp\_mr, lp\_mu, lp\_mz, m, magerr\_g, magerr\_i, magerr\_r, magerr\_u, magerr\_y, magerr\_z, mag\_g, mag\_i, mag\_lim\_g, mag\_lim\_i, mag\_lim\_r, mag\_lim\_u, mag\_lim\_y, mag\_lim\_z, mag\_r, mag\_u, mag\_y, mag\_z, mask, maxval, model-flux, mu\_max, mu\_threshold, nbpz\_filt, nbpz\_flagfilt, nbpz\_nondetfilt, n-exposures-detec, n-exposures-used, nimaflags\_iso, odds, paudm\_id, psf-e1, psf-e1-exp[1-14], psf-e2, psf-e2-exp[1-14], psf-strehl-ratio, scalelength, seqnr, snratio, star\_flag, t\_b, t\_b\_stars, theta\_j2000, t\_ml, t\_ml\_stars, weight, xpos, ypos, z\_b, z\_b\_max", z\_b\_min, z\_ml \\
\hline
\textbf{deep2} & Deep2 catalog & comment, date, dec, dof, e2, magb, magberr, magi, magierr, magr, magrerr, mask, m\_b, mjd, objname, objno, obj\_type, pa, pgal, ra, rchi2, rg, sfd\_ebv, slit, slitdec, slitlen, slitpa, slitra, star\_type, ub, vdisp, vdisperr, z, zbest, zerr, zquality \\
\hline
\textbf{gaia\_dr2} & Gaia DR2 catalog for astrometry and calibration & dec, dec\_err, duplicated\_source, phot\_bp\_mean\_flux, phot\_bp\_mean\_flux\_error, phot\_bp\_mean\_mag, phot\_g\_mean\_flux, phot\_g\_mean\_flux\_error, phot\_g\_mean\_mag, phot\_rp\_mean\_flux, phot\_rp\_mean\_flux\_error, phot\_rp\_mean\_mag, pmdec, pmdec\_error, pmra, pmra\_error, ra, gaia\_dr2, ra\_err, public, gaia\_dr2, source\_id \\
\hline
\end{tabular}
\end{table*}

\end{document}